\documentclass[superscriptaddress,aps,prb,noeprint,reprint,floatfix]{revtex4-2}
\usepackage{graphicx}
\usepackage{amssymb}
\usepackage{amsmath}
\usepackage{braket}
\usepackage{dcolumn}
\usepackage[utf8]{inputenc}
\usepackage{bm}
\usepackage{hyperref}
\usepackage{color}
\usepackage{xcolor}
\usepackage{float}

\begin{document}
\title{Ground-state phase diagram and thermodynamics of coupled trimer chains}

\author{R. R. Montenegro-Filho}
\author{E. J. P. Silva-Júnior}
\author{M. D. Coutinho-Filho}
\affiliation{Laborat\'{o}rio de F\'{i}sica Te\'{o}rica e Computacional, Departamento de F\'{i}sica, Universidade Federal de Pernambuco, 50760-901 Recife-PE, Brasil}
\date{\today}

\begin{abstract}
The density matrix renormalization group and quantum Monte Carlo method are used
to describe coupled trimer chains in a magnetic field $h$. The Hamiltonian contains exchange
terms involving the intra-trimer coupling $J_1$ (taken as the unit of energy) and the inter-trimer
coupling $J_2$, plus the Zeeman interaction for a magnetic field $h$ along the $z$ direction. Results
for the magnetization per trimer $m$ are calculated in regimes of positive and negative values
of the ratio $J = J_2 / J_1$, from which the rich field-induced ground state phase diagram $h$
\textit{versus} $J$ is derived, with the presence of Luttinger liquid, the 1/3 -- plateau ($m = 1/2$), and the one of fully polarized magnetization ($m = 3/2$). Also, the zero-field Lanczos calculation of spin-wave dispersion from the 1/3 -- plateau for $S^z = 1$ is shown at the previous regimes of $J$ values. In addition, we also report on the decay of correlation functions of
trimers along open chains, as well as the average two-magnon distribution. The ground state is ferrimagnetic 
for $0<J\leq 1$, and is a singlet for $-1\leq J<0$. In the singlet phase, the spin correlation functions along the legs present an antiferromagnetic power-law decay, similar to the spin-1/2 linear chain, thus suggesting that the ground state is made of three coupled antiferromagnetically oriented chains. In the singlet phase, the dimensionless thermal magnetic susceptibility per site normalized by $1/|J|$  gets closer to $1/\pi^2$ as the temperature $T\rightarrow 0$. For the ferrimagnetic phase, we fit the susceptibility to the experimental data for the compound 
Pb$_3$Cu$_3$(PO$_4$)$_4$ and estimate the model exchange couplings: $J_1=74.8 \text{K}$ and $J=0.4$. These values imply a range of energies for the magnon excitations that are in accord with the data 
from neutron scattering experiments on Pb$_3$Cu$_3$(PO$_4$)$_4$ for two excitation modes. The 1/3 -- plateau closes only at $1/|J|=0$ with $J<0$. 
\end{abstract}

\maketitle

\section{Introduction}

Chains of spin-1/2 sites exhibit a rich quantum behavior 
depending on the way the spins are coupled \cite{giamarchi2003quantum}. 
The spin-1/2 antiferromagnetic linear chain \cite{Muller1981} is a paradigmatic 
model with a gapless spectrum that unveiled, for example, 
collective excitations such as the spinons and holons \cite{giamarchi2003quantum}.
On the other hand, the physics of short-range valence bond states manifests 
in the singlet gapped ground state of a spin-1/2 two-leg ladder \cite{Dagotto1992,Barnes1993,Dagotto1996}, 
which has two spins in each unit cell. For ladders with $n$ legs, the ground state is gapped, 
with a finite correlation length, for $n$ even; while it is gapless, with a power-law 
decay of correlation functions, for $n$ odd \cite{White1994,Frischmuth1996}.

In a similar but more complex fashion, chains of coupled
spin-1/2 trimers have a unit cell made of three spin-1/2 sites and
exhibit rich phase diagrams. 
We notice two typical exchange pathways arrangements for unfrustrated bipartite lattices, both 
in accord with the topological Lieb-Mattis theorem \cite{LiebMattis}. In the first case \cite{Hase2006,Gu2006,Hase2007,Cheng2022,Verkholyak2021}, the lattice can be divided into two groups with the same number of sites, and the ground state is a gapless singlet. In particular, the 
magnetic features of the  compound Cu$_3$(P$_2$O$_6$OH)$_2$ are investigated through this model \cite{Hase2006,Hase2007,Hase2020}. In the second case \cite{Macedo1995,PRL97Raposo,PRB99Raposo}, there are  two unit-cell sites in 
one sublattice and one unit-cell site in the other; thus, a spontaneous break of the rotation symmetry
occurs. The ground state has a finite magnetic moment in the thermodynamic limit and
exhibits a ferrimagnetic order \cite{Tian94}. The low-energy magnon spectra \cite{PRL97Raposo,PRB99Raposo,Yamamoto2007} have a ferromagnetic gapless Goldstone mode, due to the broken symmetry, and two gapped excitations, one ferromagnetic  and the other antiferromagnetic. The Hamiltonians in this category are used to understand the magnetism of the 
ferrimagnetic phosphates A$_3$Cu$_3$(PO$_4$)$_4$ ($A=$ Ca, Sr, and Pb). 
Considering non-bipartite frustrated lattices, we mention the diamond chains \cite{Takano1996,Okamoto2003}, which have a singlet ground state and model 
the azurite compound Cu$_3$(CO$_3$)$_2$(OH)$_2$ \cite{Kikuchi2005,Rule2008,Aimo2009}.  Further, frustrated models exhibit frustration-induced magnon condensation \cite{Montenegro-Filho2008}, 
first-order phase transitions \cite{DoNascimento-Junior2019}, and pseudo phase transitions \cite{Rojas2021}.  

In any of the above-mentioned coupled trimer chains, a magnetization plateau at 1/3 of the saturation magnetization (a 1/3 -- plateau) is observed in the magnetization curve as a function of a magnetic field. This is in accord with the topological Oshikawa-Yamanaka-Affleck criteria \cite{OYAPrl97}, due to the number of unit-cell sites and the absence 
of broken translation symmetry. However, the quantum state of the three unit-cell spins depends on the parameter regime and the Hamiltonians considered. In some cases, two of the unit-cell spins are in a singlet state \cite{Okamoto2003}, while in others, the same spins are in a coherent superposition of triplet states \cite{Yamamoto2007}. Further, we mention that the edge states present at the 1/3 -- plateau phase of unfrustrated \cite{Montenegro-Filho2020} and frustrated \cite{Furuya2021} models with open boundaries were recently investigated. 

In this work, we calculate the phase diagram of coupled trimer chains by using density matrix renormalization group (DMRG) \cite{White1992,*White1993,*Schollwock2005} and quantum Monte Carlo (QMC) methods. In Sec. \ref{sec:model-and-methods}, we present the Hamiltonian of the trimer 
chain and the details of the methods mentioned above.
In Sec. \ref{sec:magnetization-and-phase-diagram}, we show the bounds of the 1/3 -- plateau and fully polarized 
plateau, besides the parameter region in which a singlet antiferromagnetic phase is found. 
In Sec. \ref{sec:low-energy-exacitations}, we discuss the low-energy excitation modes from the 1/3 -- plateau state (three magnon excitations), and from the singlet antiferromagnetic phase, a spinon excitation as in 
the spin-1/2 linear chain \cite{DesCloizeaux1962}. Further, in Sec. \ref{sec:edge-states}, we consider the edge states of open chains, which are associated with the 1/3 and the fully polarized plateaus. 
We investigate the thermal susceptibility $\chi$ at zero magnetic 
field in Sec. \ref{sec:susceptibility}, and the closing of the 1/3 -- plateau in 
Sec. \ref{sec:gapclosing}. We summarize our results  
in Sec. \ref{sec:summary}.
 
\section{Model and methods}
\label{sec:model-and-methods}
\begin{figure}[H]
\centering{\includegraphics*[width=0.36\textwidth]{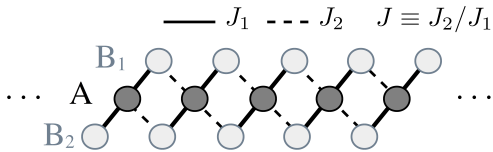}}
\caption{Scketch of the spin-1/2 trimer chain Hamiltonian.
The total number of trimers is denoted by $L$.
The intratrimer coupling is $J_1$, while the intertrimer 
one is $J_2$, the ratio between the couplings 
is defined as $J\equiv J_2/J_1$.}
\label{fig:ham}
\end{figure}

The coupled trimer chain Hamiltonian is sketched in Fig. \ref{fig:ham}. For
a chain of $L$ coupled trimers in a magnetic field, the Hamiltonian in units of the intra-trimer 
coupling $J_1$ reads  
\begin{eqnarray}
 \frac{H}{J_1} &=&\sum_{i=1}^{L}[\mathbf{A}_{i}\cdot(\mathbf{B}_{1,i}+\mathbf{B}_{2,i})\\ \nonumber 
 & &+J(\mathbf{A}_{i}\cdot\mathbf{B}_{2,i+1}+\mathbf{B}_{1,i}\cdot\mathbf{A}_{i+1})]\\ \nonumber
 & &-S^z h,
 \label{eq:hamiltonian}
\end{eqnarray}
where each trimer $i$ has three spin-1/2 spins ($\hbar\equiv 1$) at sites $A_i$, $B_{1,i}$, and $B_{2,i}$, with spin operators 
denoted by $\mathbf{A}_{i}$, $\mathbf{B}_{1,i}$, and $\mathbf{B}_{2,i}$, respectively, while the ratio $J\equiv J_2/J_1$, where 
$J_2$ is the intertrimer coupling. We define $h$ as the magnetic field in units of $J_1/g\mu_B$, where $\mu_B$ is the Bohr magneton and 
the $g$ factor is assumed the same at all sites. The total spin component in the $z$ direction (field direction) is $S^z$.
We consider $m$ as the magnetization per trimer in units of $g\mu_B$, such that $m\equiv S^z/L$.

Our zero-temperature results were obtained with the density matrix renormalization group (DMRG) on chains with open boundary conditions and exact diagonalization on closed chains.
We retained up to 243 states per block in the DMRG calculations and the maximum discarded weight was $1\times 10^{-8}$.
The thermal magnetic susceptibility was obtained with the quantum Monte Carlo (QMC) method, the looper algorithm, using the code from the Algorithms and Libraries for Physics Simulations (ALPS) project
\cite{Bauer2011}. We have used $1 \times 10^5$ Monte Carlo steps in the QMC calculations.

\section{Magnetization curves and phase diagram}
\label{sec:magnetization-and-phase-diagram}
\begin{figure}
\includegraphics*[width=0.45\textwidth]{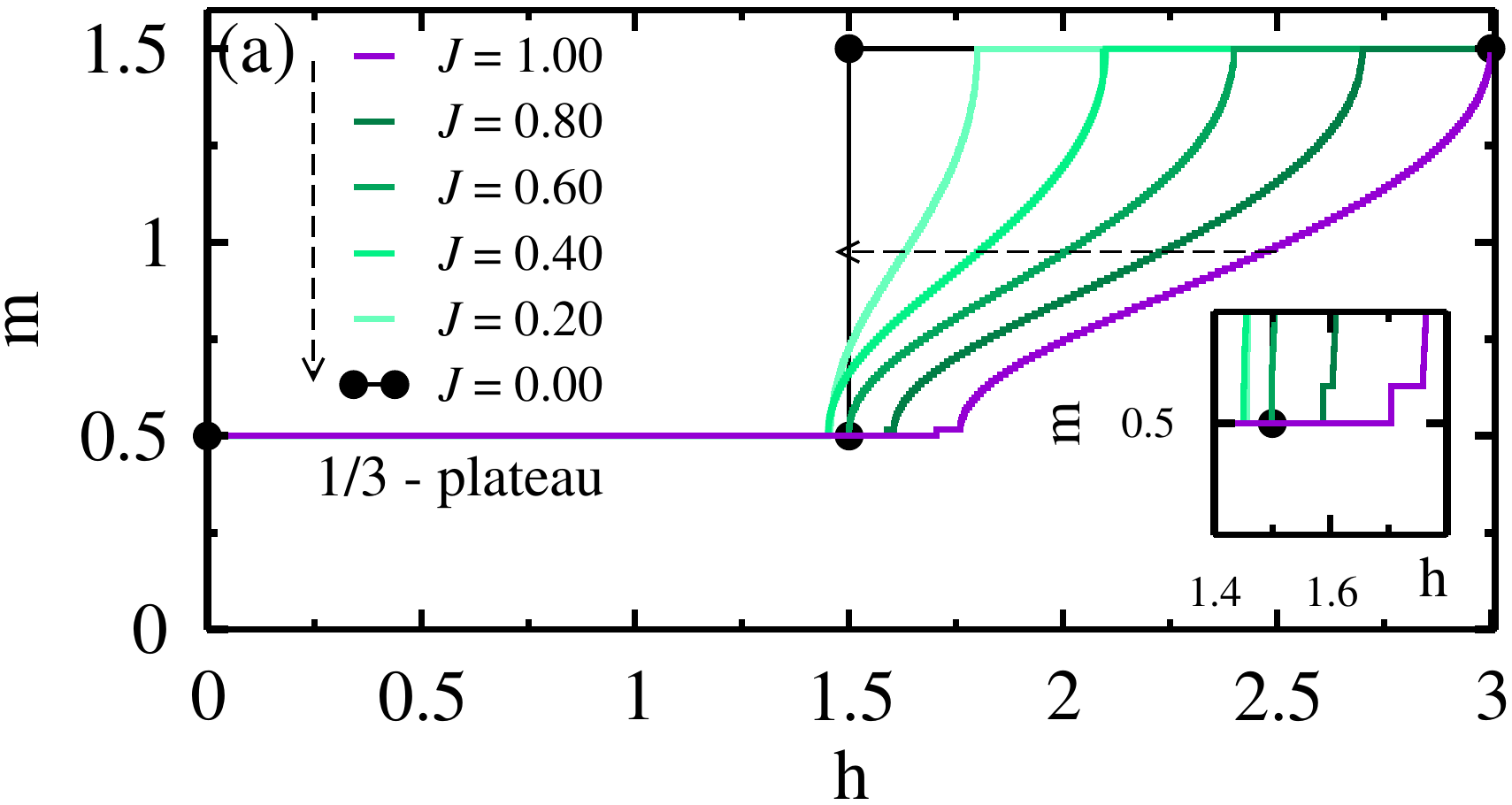}
\includegraphics*[width=0.45\textwidth]{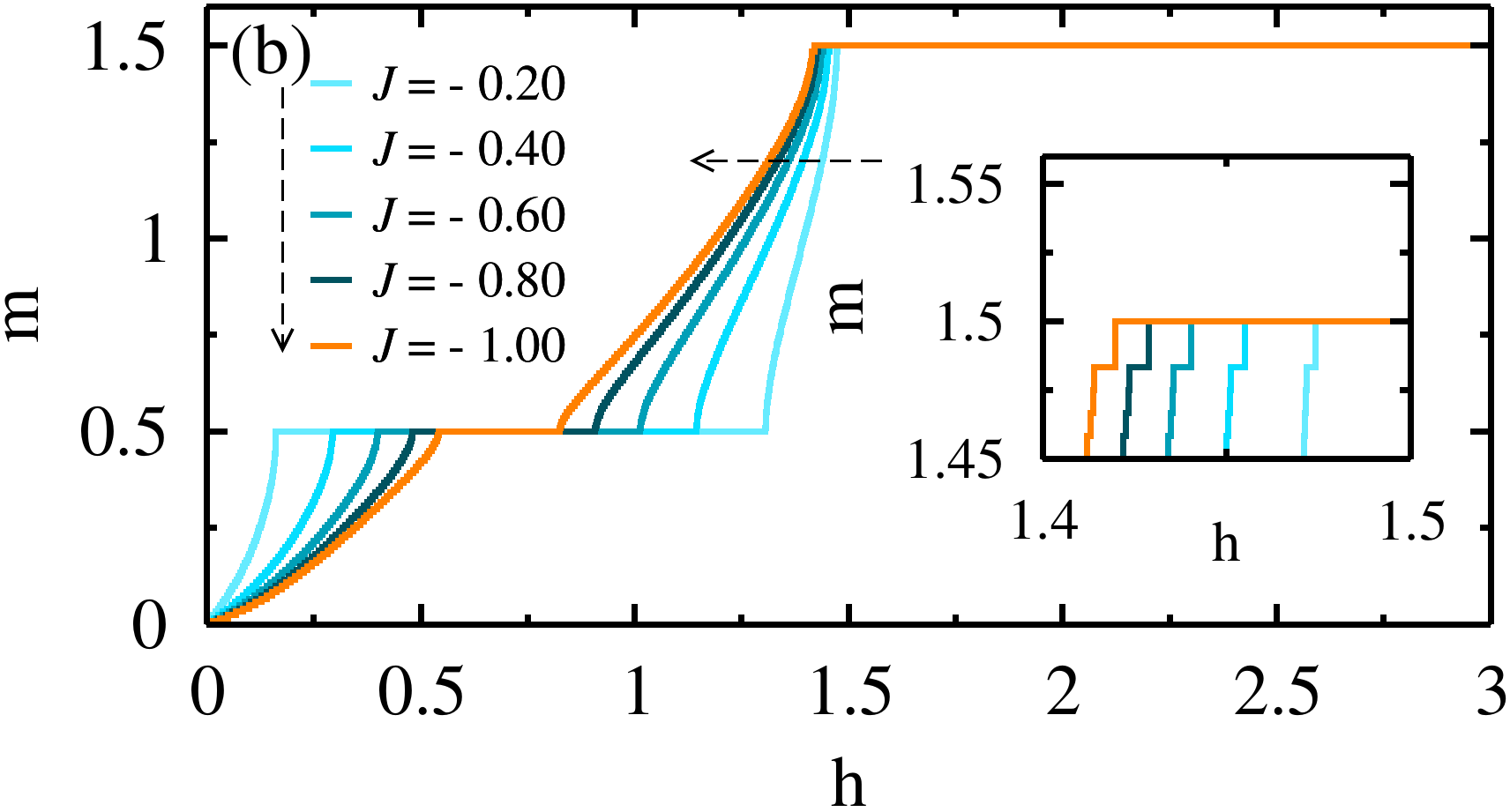}
\caption{
Magnetization per trimer $m$ as a function of the magnetic field $h$ 
for the indicated values of $J=J_2/J_1$. Density matrix renormalization 
group results for an open chain with $L=120$ trimers. The insets show a 
zoom of the curves in the vicinity of (a) the 1/3 -- plateau and (b) the 
fully polarized plateau to have a better view of the magnetizations 
of localized edge states.
}
\label{fig:mag}
\end{figure}

Considering first the case of zero field, $h=0$, the ground-state total 
spin for $J>0$ is very distinct from the $J<0$ case. For $J>0$, the ground-state total spin
is 1/2 per trimer, as shown in Fig. \ref{fig:mag}(a), and
presents a spontaneously broken rotation symmetry in spin space. On the other hand, for $J< 0$, 
the ground state is a rotationally invariant singlet state, as shown in Fig. \ref{fig:mag}(b). 
In particular, the ground-state total spin for $J>0$ is in accord with the Lieb-Mattis theorem \cite{LiebMattis},
while the trimer chain Hamiltonian does not satisfy the requirements of this theorem for $J<0$.
For $J=0$ the trimers are decoupled and there is a degeneracy between total spin states from 
0 to 1/2 per trimer. In particular, we mention that the ferrimagnetic phase is 
also observed in the phase diagram of frustrated chains, like 
the distorted diamond chain \cite{Okamoto2003}, and perturbation theories can be developed 
from the state of decoupled spin-1/2 trimers \cite{Okamoto2003}. 

\begin{figure}
\includegraphics*[width=0.45\textwidth]{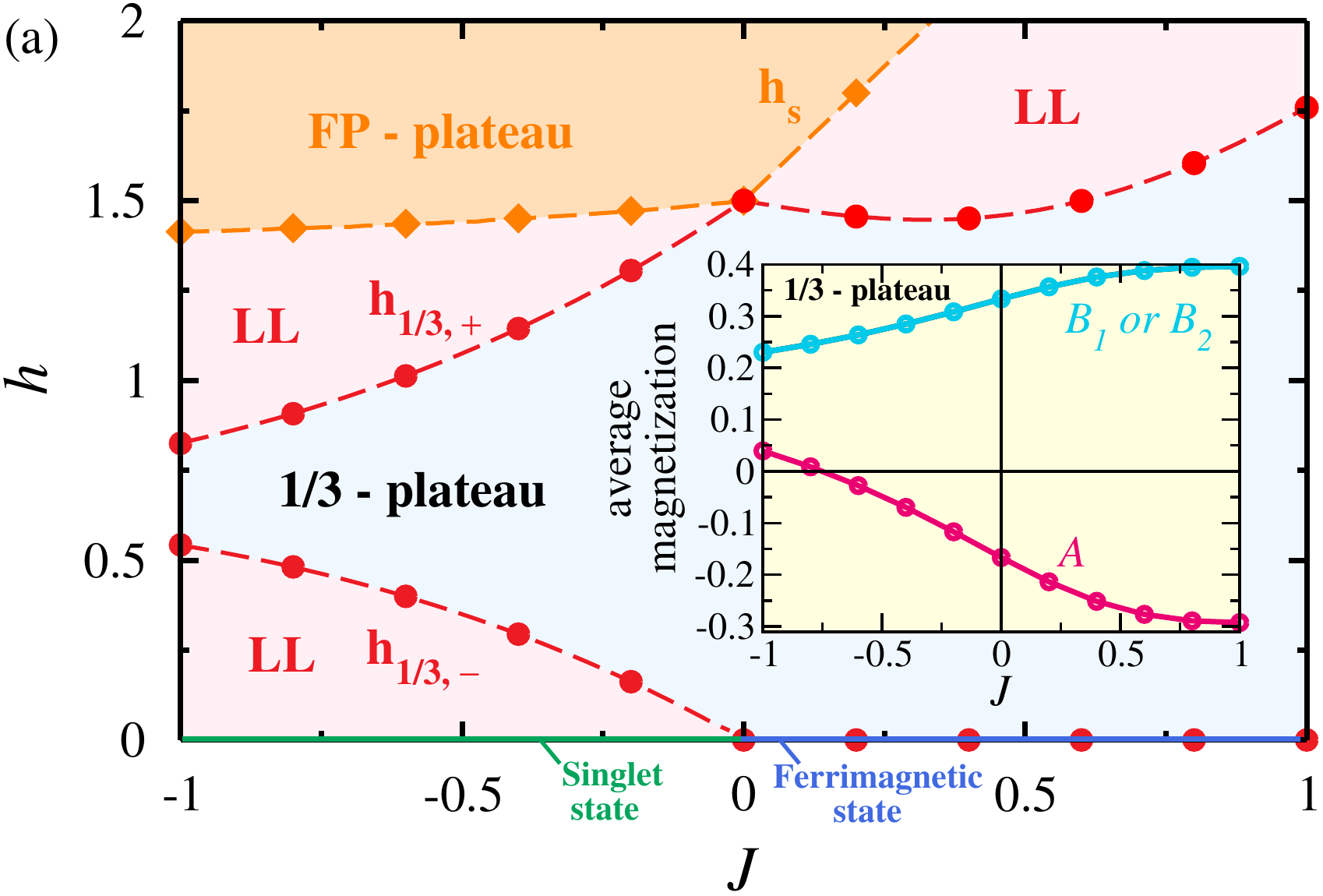}
\centering{\includegraphics*[width=0.42\textwidth]{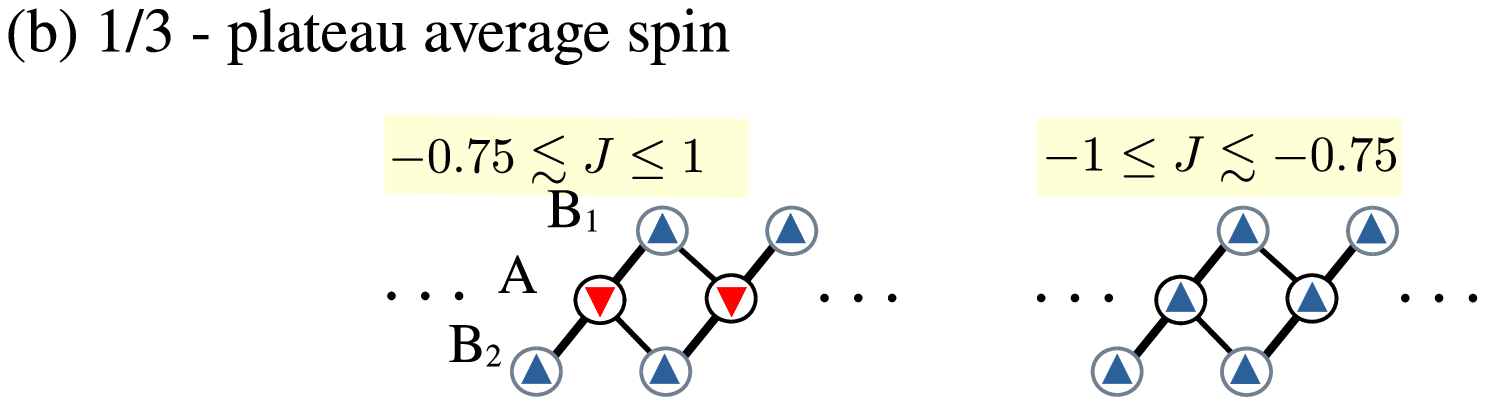}}
\caption{(a) Magnetic field $h$ vs. exchange ratio $J$ phase diagram of the trimer chain.
We show the transition lines from the gapped 1/3 -- plateau, $h_{1/3,+}$ and 
$h_{1/3,-}$, and the fully polarized (FP) plateau, $h_s$, to gapless Luttinger liquid (LL) phases. The phase diagram is estimated from the DMRG results 
for chains with up to $L=120$ trimers. For $h=0$, the ground state is a singlet for 
$-1\leq J<0$, and is ferrimagnetic for $0<J\leq 1$. In the inset we present 
the average spin at $A$ and $B_1$ (or $B_2$) sites at the 1/3 -- plateau. 
The relative orientation of the average spins changes from a ferrimagnetic pattern, found
in the range $-0.75<J<1$, to a ferromagnetic orientation in $-1<J<-0.75$, as sketched in (b). 
}
\label{fig:diag}
\end{figure}

In the presence of a magnetic field $h\neq 0$, the magnetization curves of the trimer chain, 
for $J\geq 0$ [Fig. \ref{fig:mag}(a)] and $J<0$ [Fig. \ref{fig:mag}(b)], present two magnetization plateaus: 
the fully polarized plateau (FP plateau) at the saturation magnetization $m_s=3/2$ and the
1/3 -- plateau at magnetization $m=m_s/3=1/2$. These are integer magnetization plateaus, they 
do not break the translation symmetry of the Hamiltonian, and are in accord with the 
topological Oshikawa-Yamanaka-Affleck criteria \cite{OYAPrl97}. In the insets of Figs. \ref{fig:mag}(a) and (b) we 
show the presence of localized edge states, manifested as in-gap magnetization states inside the 
thermodynamic-limit plateaus. The edge states are associated with the topological nature of the magnetization plateaus. 
We notice the presence of edge states related to the 1/3 -- plateau in a region of positive values of $J$ and to the
FP--plateau for $J<0$. 

We present in Fig. \ref{fig:diag}(a) the phase diagram of the trimer chain as a function of magnetic field  
and of the ratio $J$. The gapped FP--plateau is bounded by the saturation field $h_s$, while the 1/3 -- plateau phase is bounded 
by the lower and higher critical fields, $h_{1/3,-}$ and $h_{1/3,+}$, respectively. We notice that $h_{1/3,-}=0$ 
for $J\geq 0$, since, in zero field, the ground state total spin is 1/2 per trimer and is degenerate for $J=0$.
The gapped phases are separated by gapless Luttinger liquid (LL) phases. We also show the average spin 
at $A$ and $B$ sites in the inset of Fig. \ref{fig:diag}(a) for the 1/3 -- plateau magnetization. 
The average spin in a trimer presents an antiferrromagnetic arrangement for $-0.75\lesssim J\leq 1$, 
implying a ferrimagnetic ground state; and a ferromagnetic orientation for $-1\leq J\lesssim -0.75$, 
as sketched in Fig. \ref{fig:diag}(b). 

\section{Low energy excitations}
\label{sec:low-energy-exacitations}
In this section we discuss the low-energy magnetic excitations from 
the 1/3 -- plateau phase and from the singlet ground state of 
the $-1\leq J<0$ region. We obtain the energy $E_q(S^z)$ 
of finite chains with periodic boundary conditions as a function 
of the lattice wave vector $q$ for a fixed value of $S^z$ 
by using exact diagonalization.

\begin{figure}
\includegraphics*[width=0.49\textwidth]{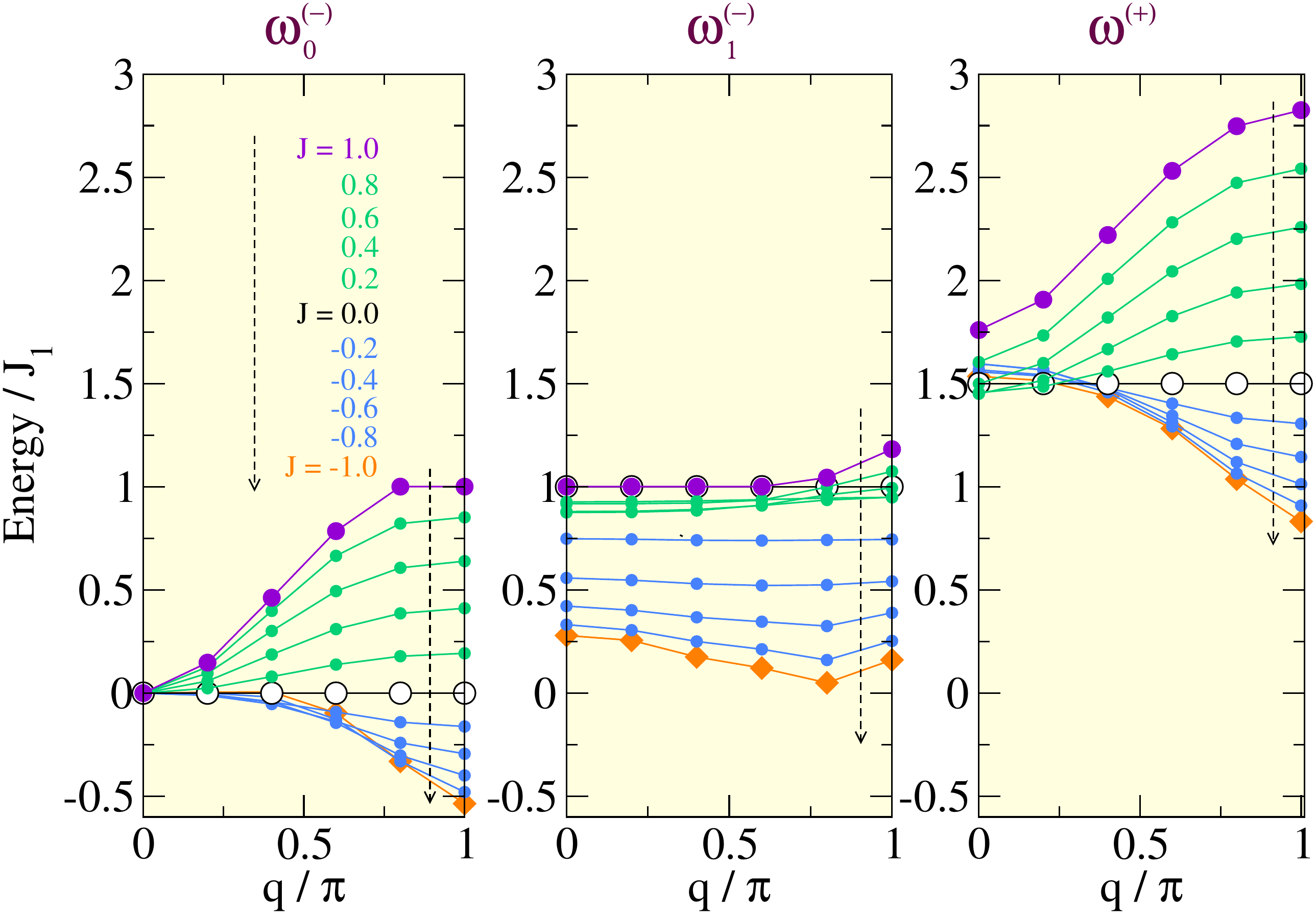}
\caption{Spin-wave modes from the 1/3 - plateau magnetization ($m=1/2$) for
the indicated values of $J$ and magnetic field $h=0$, where $q$ is the lattice 
wave vector. The modes $\omega_0^{(-)}$ and $\omega_1^{(-)}$
have spin $\Delta S^z=-1$, while the mode $\omega^{(+)}$ has spin $\Delta S^z=1$.
The data are exact-diagonalization 
results for a closed chain with $L=10$ trimers.
}
\label{fig:espectro13}
\end{figure}

For the 1/3 -- plateau state, we consider three magnetic excitations 
in zero magnetic field, 
as shown in Fig. \ref{fig:espectro13}. 
Two of them, $\omega^{(-)}_0$ and $\omega^{(-)}_1$, carry a spin $\Delta S^z=-1$
and are given by
\begin{equation}
\omega^{(-)}_n(q)=E_q^{n}(S^z=L/2-1)-E_{q=0}(S^z=L/2),
\end{equation}
for $n=0$, lowest energy in the sector $(q,S^z=S_{GS}-1)$, and 
$n=1$, first excitation in the same sector. Notice that 
the ground state in zero field has the total spin 
$S_{GS}=L/2$ for $J\geq0$; while in the range $-1\leq J\leq 0$, 
this is the total spin of the ground state only 
for $h_{1/3,-}\leq h \leq h_{1/3,+}$. In fact, as shown in 
Fig. \ref{fig:espectro13}, $\omega^{(-)}_0(q)<0$ in zero field 
for $J<0$, with the $q=\pi$ sector as the lowest one. 
For $h\neq 0$, we add the Zeeman term, $-hS^z$, to the energy  
and the excitation energy changes as 
\begin{equation}
\omega^{(-)}_0(q,h)=\omega^{(-)}_0(q)+h.
\end{equation}
Hence, the critical field $(h=h_{1/3,-})$ for which the ground state has $S^z=L/2$ 
can be obtained from the condition $\omega^{(-)}_0(q=\pi,h_{1/3,-})=0$, such that 
$h_{1/3,-}=-\omega^{(-)}_0(\pi)$. The values of $h_{1/3,-}$ thus obtained are in 
excellent accord with those from DMRG, shown in 
Figs. \ref{fig:mag} and \ref{fig:diag}.

The excitation $\omega^{(-)}_1(q)$ is higher than $\omega^{(-)}_0(q)$, 
approximately dispersionless and near 1 for $J>0$. In fact, for $J=1$
the chain has a local parity symmetry \cite{PhysA2005} and an exact dispersionless 
mode in the sector $S^z=L/2-1$ with one singlet between $B_2$ of one trimer and the 
$B_1$ of the nearest neighbor trimer. In fact, there is a crossing between localized and 
dispersive modes as follows. The localized excitation is observed \cite{PhysA2005} in the 
last two points near the zone boundary of the excitation $\omega^{(-)}_0(q)$, while 
the last two points near the zone boundary of the excitation $\omega^{(-)}_1(q)$ are, 
in fact, a continuation of the dispersive mode observed in the four points near 
$q=0$ of $\omega^{(-)}_0(q)$. As shown in Fig. \ref{fig:espectro13}, for $0<J<1$ 
the excitation $\omega^{(-)}_1(q)$ gains some dispersion, implying a certain mobility of 
the local singlet, but there is not a crossing 
between the modes, since in this case there is not a local symmetry.´Also, 
$\omega^{(-)}_1(q)$ acquires a minimum in a value of $q\neq0\text{ and }\pi$.

The third excitation in 
Fig. \ref{fig:espectro13}, $\omega^{(+)}(q)$, carries a spin
$\Delta S^z=+1$ and is calculated from
\begin{equation}
\omega^{(+)}(q)=E_q(S^z=L/2+1)-E_{q=0}(S^z=L/2).
\end{equation}
For $J>0$ the minimum energy is observed at $q=0$, while for $-1<J<0$ the 
minimum is found at $q=\pi$. In the presence of a magnetic field, this 
mode condenses in the critical field $h=h_{1/3,+}$. Since, for $h\neq0$, 
we have:
\begin{equation}
\omega^{(+)}(q,h)=\omega^{(+)}(q)-h, 
\end{equation}
the condition at critical field is $\omega^{(+)}(q_{min},h=h_{1/3,+})=0$, where $q_{min}=0$ for 
$J>0$ and $q_{min}=\pi$ for $J<0$. The critical field thus obtained is 
also in excellent agreement with the DMRG results, shown in 
Figs. \ref{fig:mag} and \ref{fig:diag}. For $0\leq J\leq 1$, we notice that analytical results \cite{Yamamoto2007} from perturbation theory or interacting spin-wave analysis are in good accord with the exact diagonalization data depending on the value of $J$. Perturbation theory is better for lower values of $J$, $J\lesssim 0.5$, while spin-wave analysis gives better results for $0.5\lesssim J\leq 1$.

We start the discussion of the low-energy magnetic excitations from 
the singlet ground state of the trimer chain in zero field, for 
$-1<J<0$, by considering the spin correlation functions in the 
ground state. We use DMRG in open chains to calculate the correlation 
functions in sublattices $A$ and $B$ as follows:
\begin{equation}
 C_X(l)=\langle \langle \mathbf{S}_{X,i}\cdot \mathbf{S}_{X,j} \rangle\rangle_{|i-j|=l}, 
\end{equation}
where $\langle \ldots \rangle_{|i-j|=l}$ is an average
over all pairs of trimers $i$ and $j$ separated by the distance $l$, 
$X=A\text{ or } B$, and $B$ can be either $B_1$ or $B_2$, since 
the correlations are equal along the sublattices $B_1$ and $B_2$.
\begin{figure}
\includegraphics*[width=0.49\textwidth]{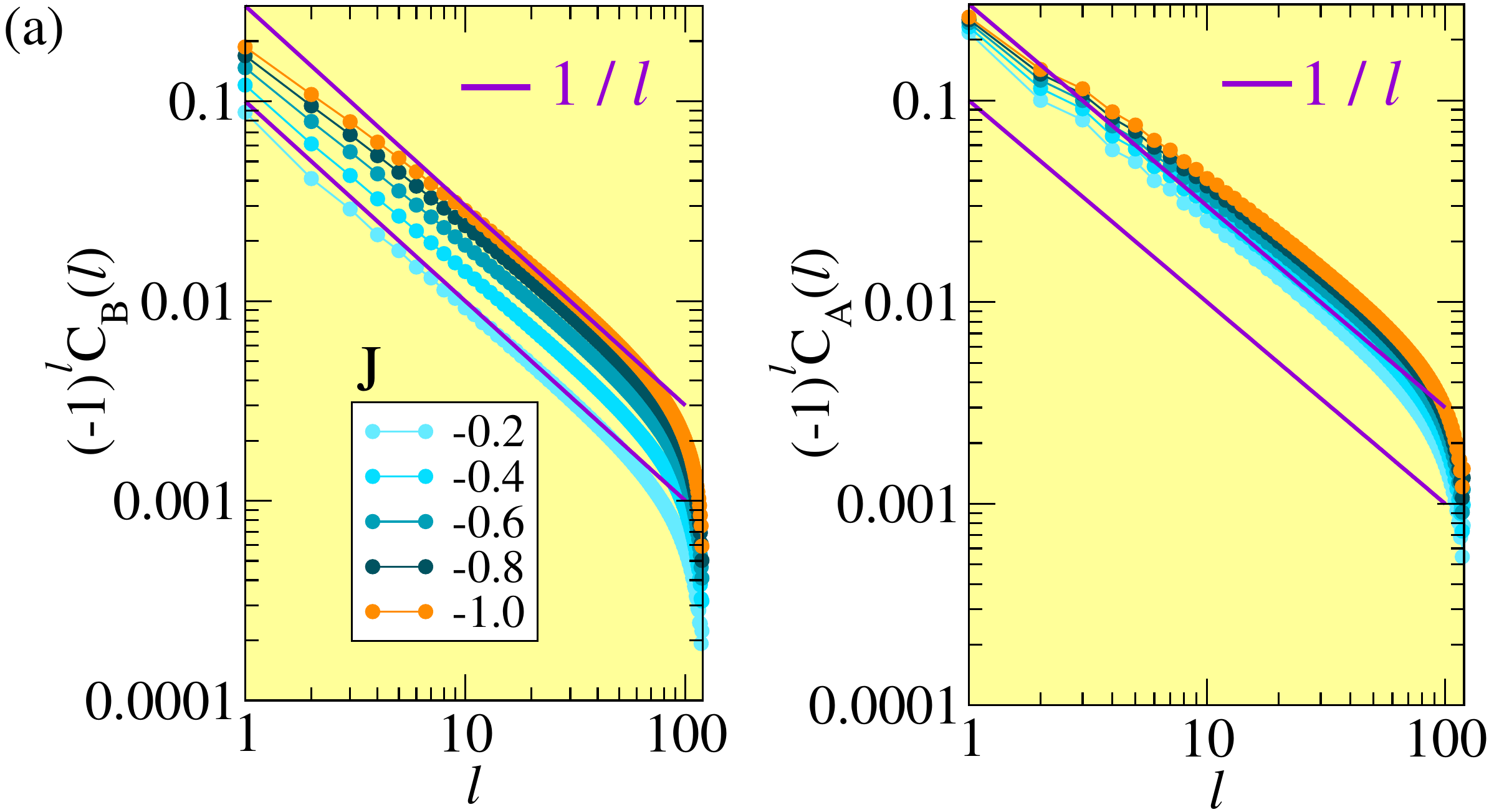}
\includegraphics*[width=0.43\textwidth]{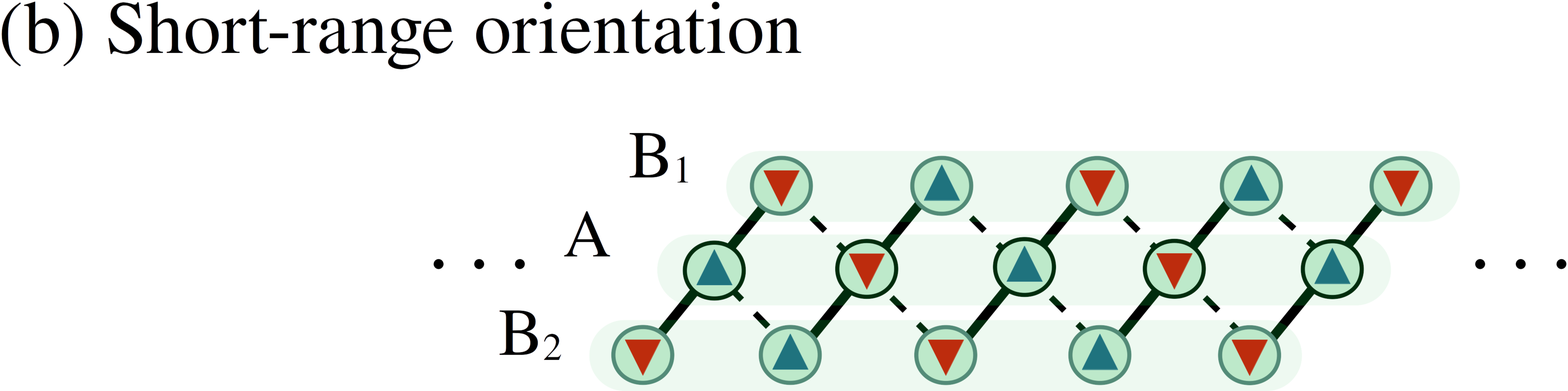}
\includegraphics*[width=0.43\textwidth]{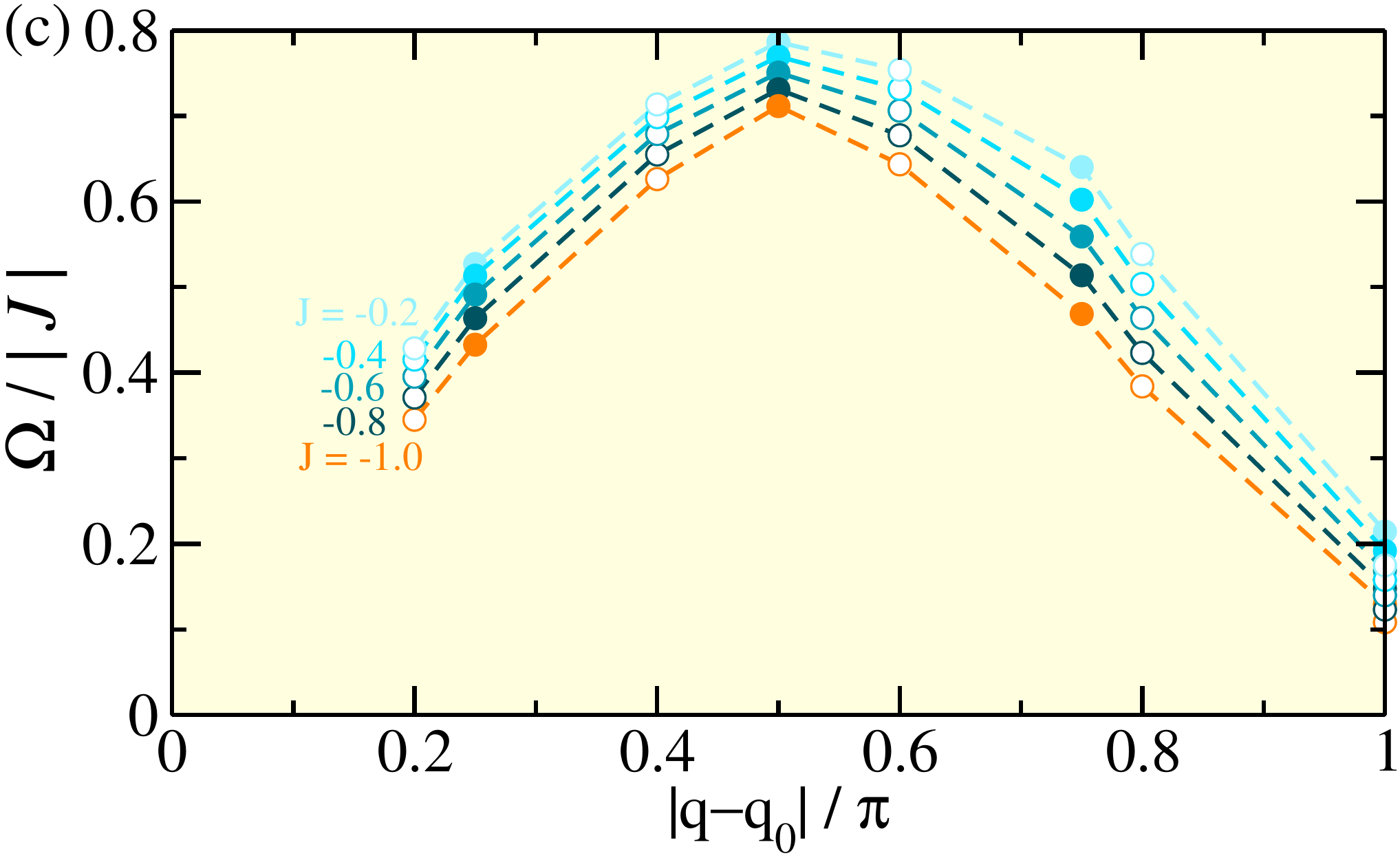}
\caption{
(a) Density matrix renormalization results for the alternating correlation functions between spins at $B_1$ (or $B_2$) sites and at $A$ sites, $(-1)^lC_B(l)$ (left panel) and $(-1)^lC_A(l)$ (right panel), respectively, as a function of the 
distance $l$ between trimers, for $J<0$ and the $J$ values indicated 
in the figure, for an open chain with $L=120$ trimers. The intermediate-distance behavior 
suggests a $1/l$ (full lines) antiferromagnetic power-law correlation along each sublattice, $A$, $B_1$ or $B_2$, 
as sketched in (b). 
(c) Exact-diagonalization results for the low-lying magnetic excitation mode 
from $m=0$ , with total spin $S=1$, for $J=J_2/J_1=-1.0,-0.8,-0.6,-0.4,\text{ and }-0.2$ (from bottom to top) as 
a function of a translated lattice wave vector $|q-q_0|$ in a closed chain, normalized by 
$|J|$. 
We present the data for two system sizes $L=8$ (filled symbols) and $L=10$ (open symbols). 
The ground state energy (with $S=0$) is observed at the wave vector $q=q_0$ with
$q_0=\pi$ for $L=8$ and $q_0=0$ for $L=10$.
}
\label{fig:espectro0}
\end{figure}

The spin correlation functions are shown in Fig. \ref{fig:espectro0}(a)
and exhibit an alternating pattern. Also, for sufficiently low distances to avoid 
edge effects, the correlation functions have an $1/l$ decay. In a linear antiferromagnetic
chain \cite{giamarchi2003quantum}, the asymptotic behavior of the spin correlation functions in 
the singlet ground state, at zero magnetic field, has the following form:
\begin{equation}
 \langle \mathbf{S}(0)\cdot\mathbf{S}(l)\rangle\sim c_1\frac{1}{l^2}+c_2(-1)^l\frac{\ln^{1/2}(l)}{l},
\label{eq:af}
\end{equation}
where the amplitudes $c_1$ and $c_2$ are not universal.
At large distances, the dominant term in Eq. (\ref{eq:af}) is $(-1)^x/l$, discarding the $\ln$ factor.
Thus, in the trimer chain, our data suggests that the spins in each sublattice 
are correlated as in the antiferromagnetic spin-1/2 linear chain, following 
the form in Eq. (\ref{eq:af}), with a short-range magnetic 
pattern as sketched in Fig.\ref{fig:espectro0}(b).  

The low-energy magnetic excitations from the ground state, $\Omega$, shown in Fig. \ref{fig:espectro0}(c) reinforce this 
picture. In particular, the antiferromagnetic spin-1/2 linear chain presents an alternation of 
the ground-state wave vector between the values $0$ and $\pi$, as shown from the Bethe ansatz solution 
\cite{DesCloizeaux1962}. On the other hand, the low-energy excitations of the 
spin-1/2 linear chain are spinons \cite{giamarchi2003quantum}, which, in the thermodynamic limit, 
have the dispersion relation \cite{DesCloizeaux1962}
\begin{equation}
 \varepsilon(q)^{(\text{spinon})}=\frac{\pi}{2}J_{\text{linear chain}}|\sin(q)|,
 \label{spinon}
\end{equation}
where $J_{\text{linear chain}}$ is the superexchange coupling between nearest
neighbor spins.

We use exact diagonalization to calculate $\Omega(q)$ for finite chains with 
periodic boundary conditions for $-1\leq J<0$ from 
\begin{equation}
 \Omega(q)=E_q(S^z=1)-E_{q=q_0}(S^z=0),
\end{equation}
where $q_0$ is the ground-state wave vector, which is $q_0=\pi$ if $L$ 
is a multiple of 4 and $q_0=0$ otherwise, for $L$ even, as 
in the antiferromagnetic spin-1/2 linear chain. In Fig. \ref{fig:espectro0}(c) 
we show $\Omega(q)$ for $L=8$ and $L=10$ as a function of $|q-q_0|$, normalized 
by $|J|$. Although the system sizes are short, we can suggest 
some features for the thermodynamic limit. The general trend of the curves 
points to a periodicity like in Eq. (\ref{spinon}), but the 
dependence on $J$ is not as simple 
as in Eq. (\ref{spinon}) for the system sizes shown. 

\section{Edge states}
\label{sec:edge-states}
The 1/3 and the fully polarized plateaus satisfy the 
Oshikawa-Yamanaka-Affleck (OYA) topological criterion for 
their occurrence in magnetization curves of 
magnetic insulators \cite{OYAPrl97}: $(m_u-S_u)=\text{integer}$,
where $m_u$ and $S_u$ are respectively 
the magnetization and the maximum spin 
in a unit period of the ground-state wave-function. In the trimer chain, there is 
not a broken translation symmetry; thus $m_u=1/2$ for the 1/3 -- plateau and 
$m_u=3/2$ for the fully polarized state, with $S_u=3/2$ in both cases. 
Further, the non-trivial topological nature of a magnetic state 
gives rise to zero-energy edge states, as expected from the 
bulk-edge correspondence \cite{Hu2014,Hu2015}. In fact, edge-mode properties have proved useful
in investigating topological quantum phase transitions 
\cite{Griffith2018,Rufo2019,MartinezAlvarez2019, Rufo2021,Montenegro-Filho2020}. 
On the other hand, it was shown \cite{Watanabe2021} that spin-$S$ antiferromagnets on $d$-dimensional 
lattices present fractionally quantized magnetization at the corners, similarly to ionic
crystals. In addition, the edge magnetization can have a non-topological origin \cite{Furuya2021}. 
In particular, in Ref. \onlinecite{Furuya2021} the authors have shown that $U(1)$ and 
site-inversion symmetry protects the edge 
magnetizations in the ferrimagnetic phase of the Union Jack stripe.
Besides, the appearance of magnon densities at the edges of alternating ferrimagnetic chains 
depends on magnon interactions \cite{DaSilva2021} and 
the uniform local potentials of the associated Holstein-Primakoff Hamiltonians.     
\begin{figure}
\begin{center}
\includegraphics*[width=0.43\textwidth]{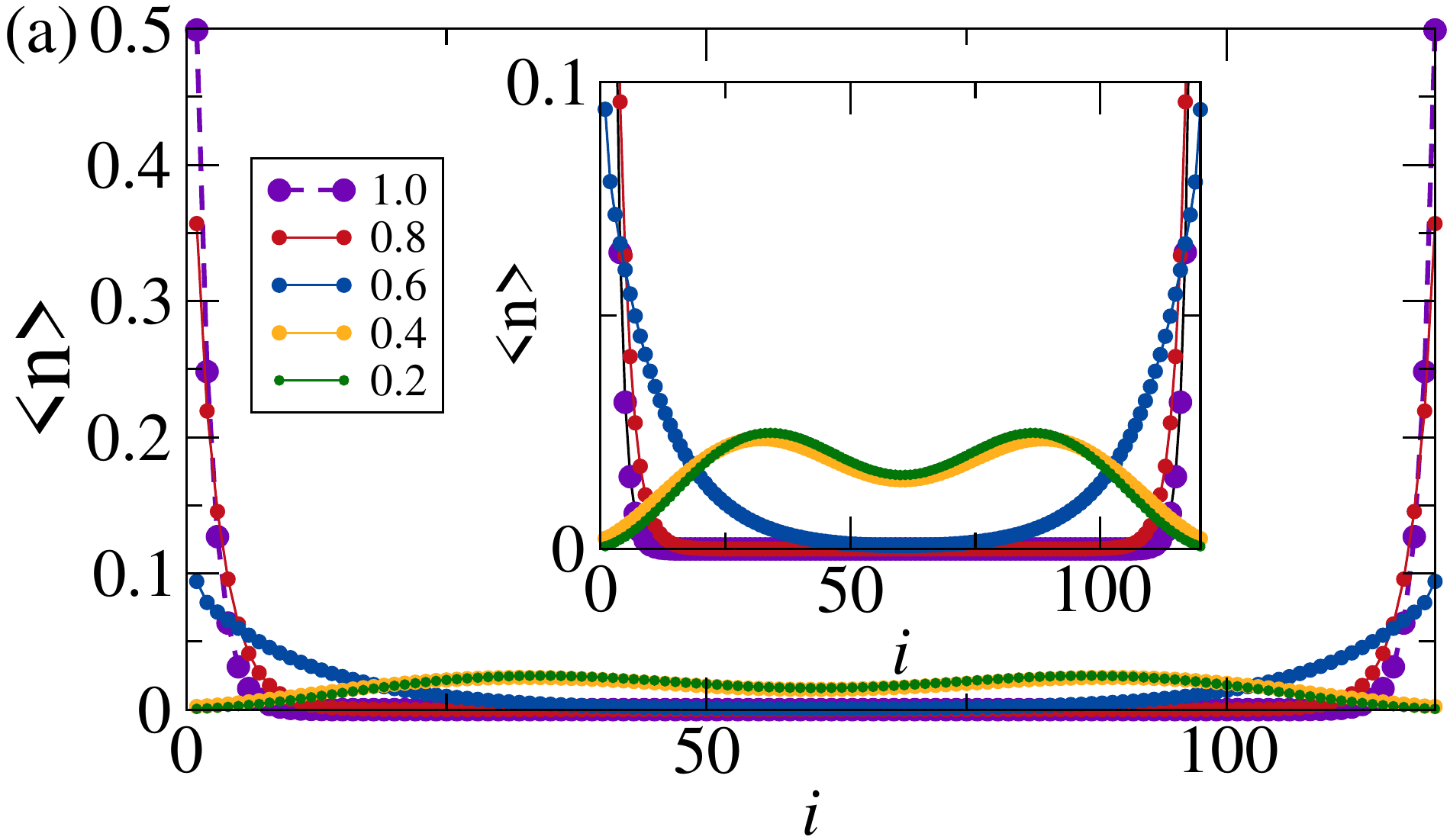}
\includegraphics*[width=0.43\textwidth]{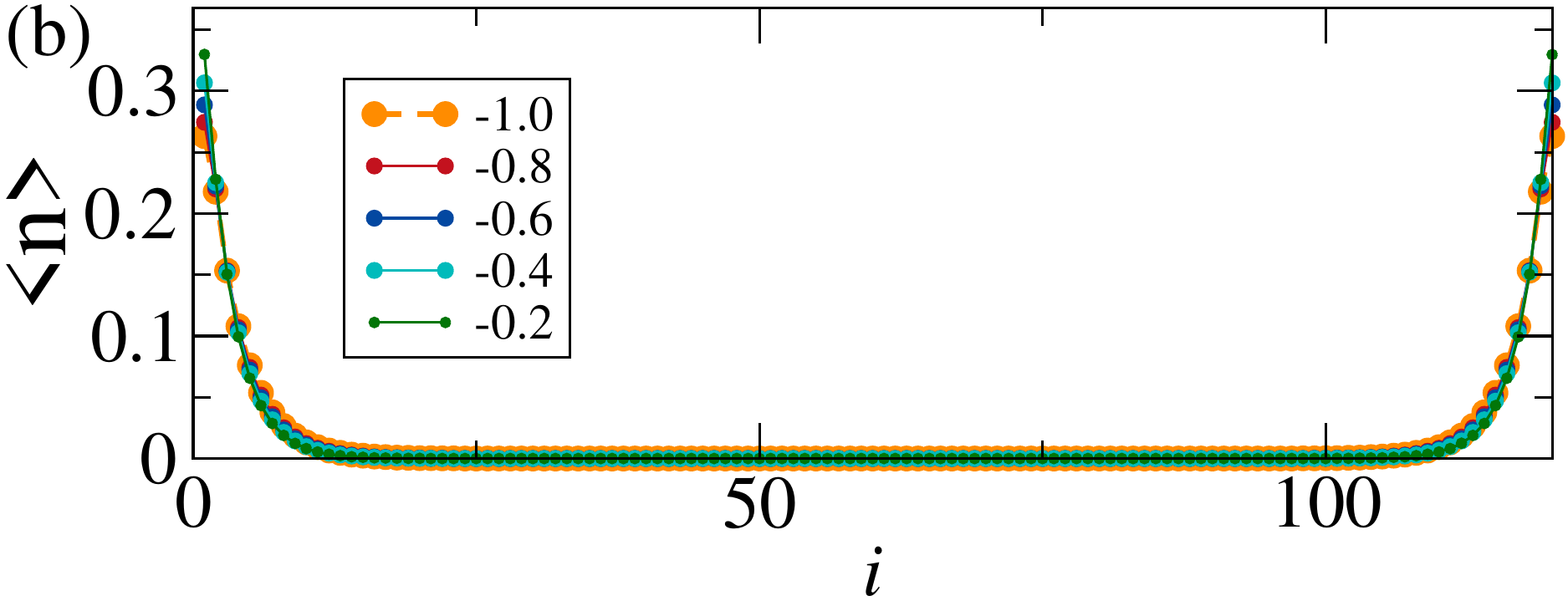}
\end{center}
\caption{Average magnon distribution in the state with two magnons added to (a) 1/3 -- plateau:
$\langle n\rangle=\langle S^z_l\rangle_{S^z=(L/2)+2}-\langle S^z_l\rangle_{S^z=L/2}$ (the inset is a zoom of the main figure), and (b) the FP-plateau: $\langle n\rangle=\langle S^z_l\rangle_{S^z=3L/2}-\langle S^z_l\rangle_{S^z=3L/2-2}$, 
with $l$ as the trimer index, for an open chain with $L=120$ trimers.}
\label{fig:edge}
\end{figure}

In Fig. \ref{fig:edge}, we present the edge magnon densities  
of the trimer chain by considering the magnon distribution as two magnons are added 
to the insulating states at $m=1/2$ (ferrimagnetic state) and $m=3/2$ (fully polarized state). 
As evidenced in the inset of Figs. \ref{fig:mag}(a) and \ref{fig:mag}(b), 
the localized magnetization states inside the thermodynamic plateaus appear for $m=3/2$ and 
$J<0$, and for $m=1/2$ and $J>0$. 
The magnons carry a spin $\Delta S^z=+1$ in the case of 
$m=1/2$, and a spin $\Delta S^z=-1$ for $m=3/2$, such that 
their distribution along the chain are calculated 
through
\begin{equation}
\langle n\rangle=\langle S^z_l\rangle_{S^z=(L/2)+2}-\langle S^z_l\rangle_{S^z=L/2}
\end{equation}
for $m=1/2$, and through
\begin{equation}
\langle n\rangle=\langle S^z_l\rangle_{S^z=3L/2}-\langle S^z_l\rangle_{S^z=3L/2-2},
\end{equation}
for $m=3/2$, where $S^z_l$ is the spin of the trimer $l$.

For the ferrimagnetic state at $m=1/2$, we observe in Fig. \ref{fig:edge}(a) the presence of magnon density at the edges 
of the trimer chain for $J>0$. However, the magnon distribution becomes less 
localized as $J$ decreases from $J=1$, and turns into bulk 
excitations for $J<0.5$. As in the case of ferrimagnetic alternating 
chains \cite{DaSilva2021}, the magnon densities at the edges can have two contributions: 
one from the geometry of the system and the other from the magnon-magnon interactions. 
In the case of the fully polarized plateau, the edge magnon density shows up for $J<0$. 
Here, the sites at the 
edges are coupled to the chain by only one antiferromagnetic bonding, 
while the other sites have at least one ferromagnetic bond: $J<0$.
Thus, the localization of the two magnons at 
the edges of the chain minimizes the total energy, and the edge magnon densities have only a geometrical origin.
\begin{figure}
\centering{\includegraphics*[width=0.45\textwidth]{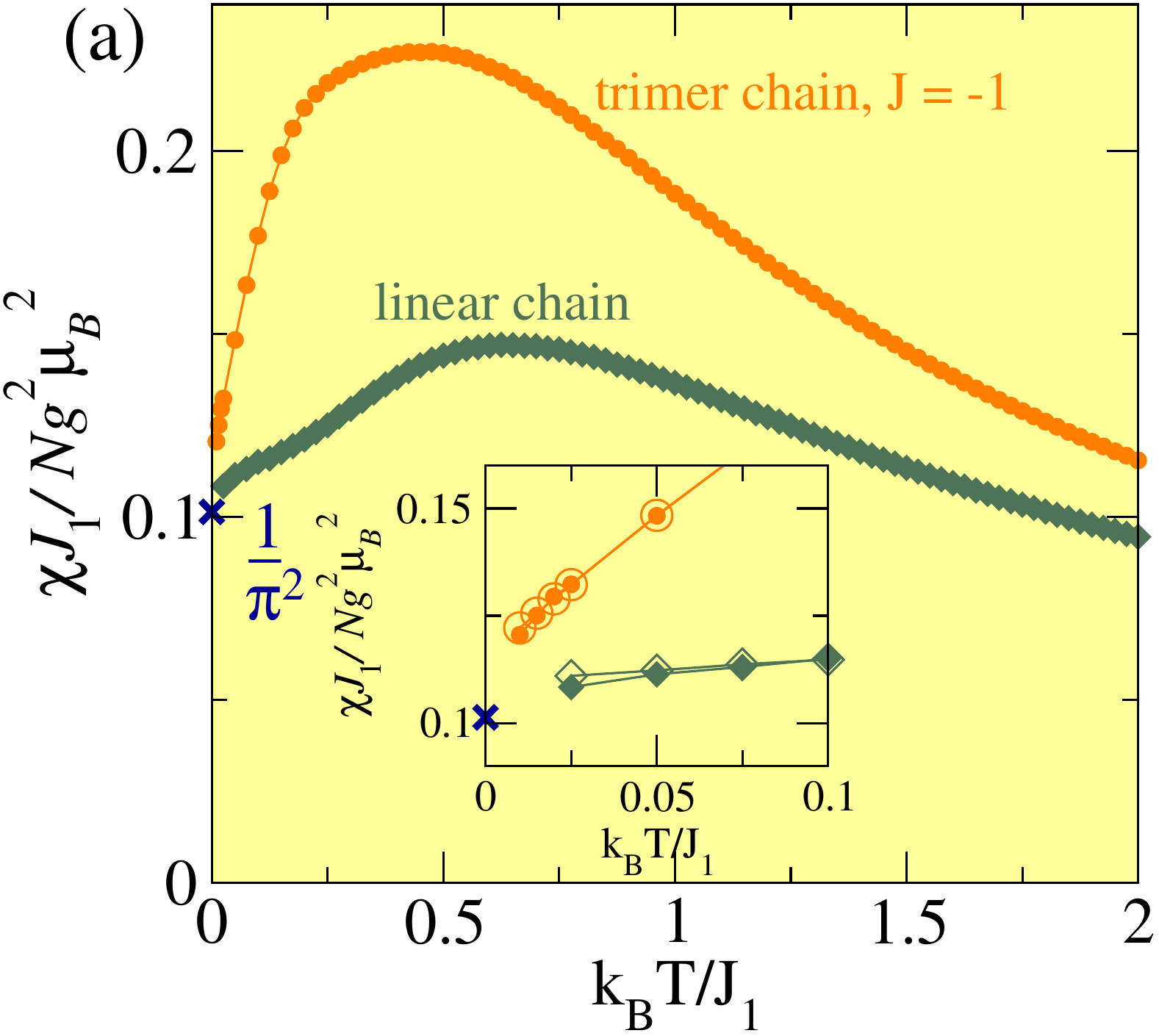}}
\centering{\includegraphics*[width=0.45\textwidth]{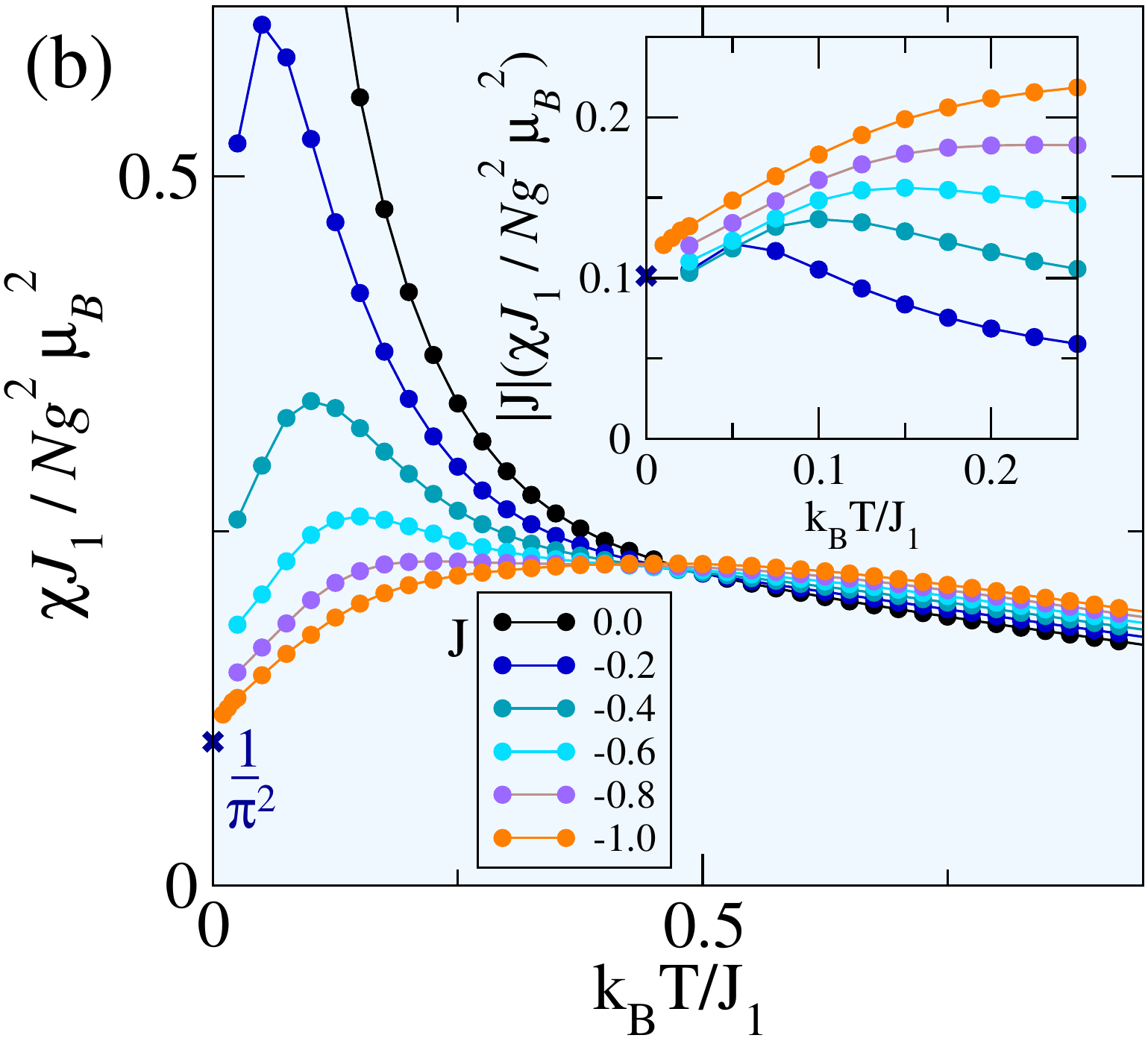}}
\caption{Quantum Monte Carlo results for the magnetic susceptibility $\chi$ per site. 
(a) Trimer chain with $L=128$ and $J=-1$, and antiferromagnetic 
linear chain with $N=128$ sites. The cross indicates the exact value of $\chi$ at 
$T=0$ for the antiferromagnetic linear chain: $1/\pi^2$. In the inset we highlight
low-temperature regime. Open symbols are results for a trimer chain with $L=256$ and 
for the antiferromagnetic linear chain with $N=256$. (b) Trimer chain for the indicated values of $J$, the cross also marks the value $1/\pi^2$. In the inset we present $|J|\chi J_1$
which is equal to $|J_2|\chi$.}
\label{fig:suscepTneg}
\end{figure}

\section{Susceptibility}
\label{sec:susceptibility}
This section discusses the finite-temperature magnetic susceptibility of the
trimer chain with quantum Monte Carlo simulations. Initially, we examine the antiferromagnetic regime, $J<0$, and, next, the
ferrimagnetic phase, observed for $J>0$. In the last case, we compare our results with the experimental
data for the Pb$_3$Cu$_3$(PO$_4$)$_4$ phosphate \cite{Belik2005,Matsuda2005}, 
which is a ferrimagnetic quasi-one-dimensional compound.
\begin{figure}
\includegraphics*[width=0.46\textwidth]{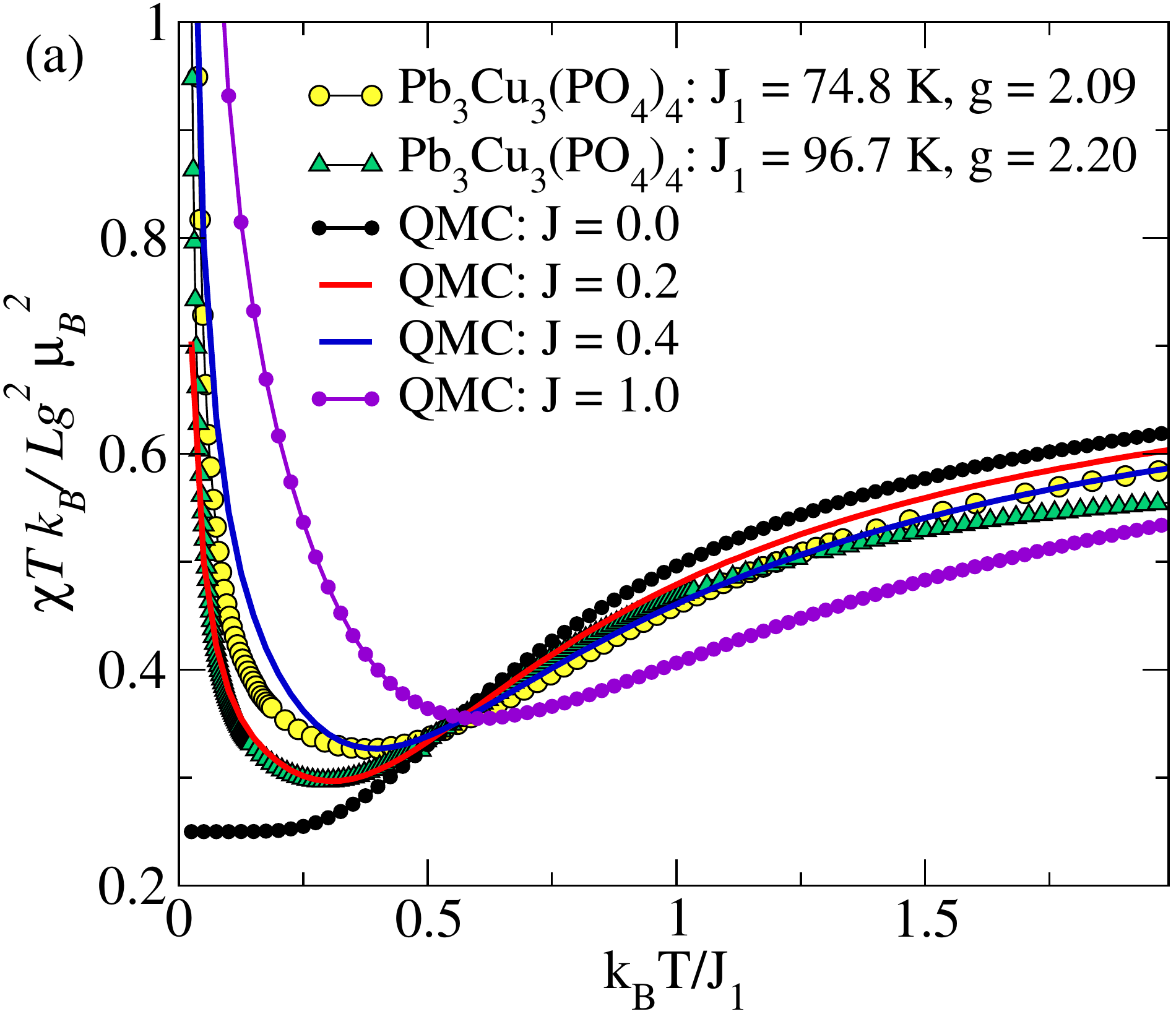}
\centering{\includegraphics*[width=0.46\textwidth]{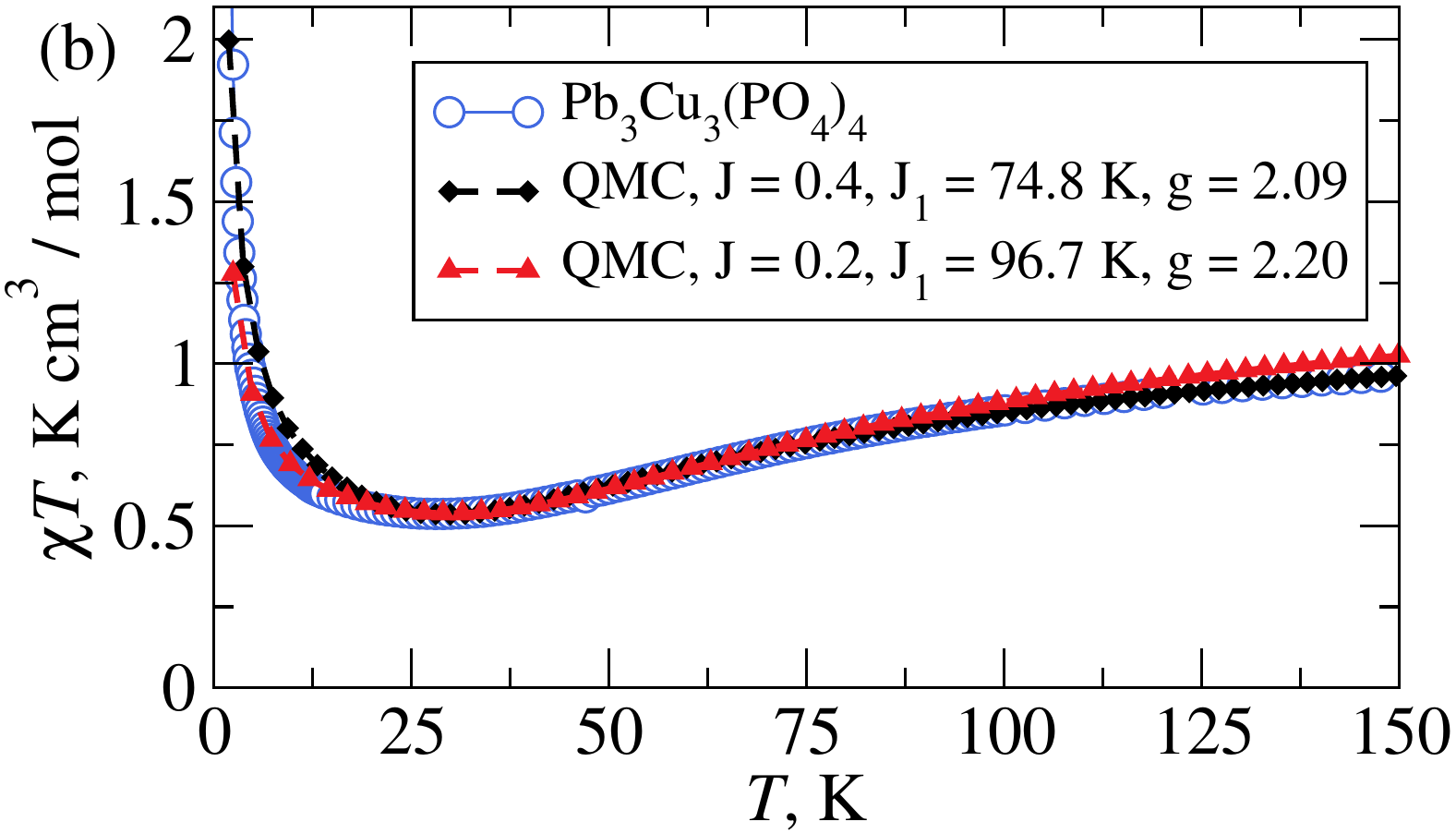}}
\includegraphics*[width=0.42\textwidth]{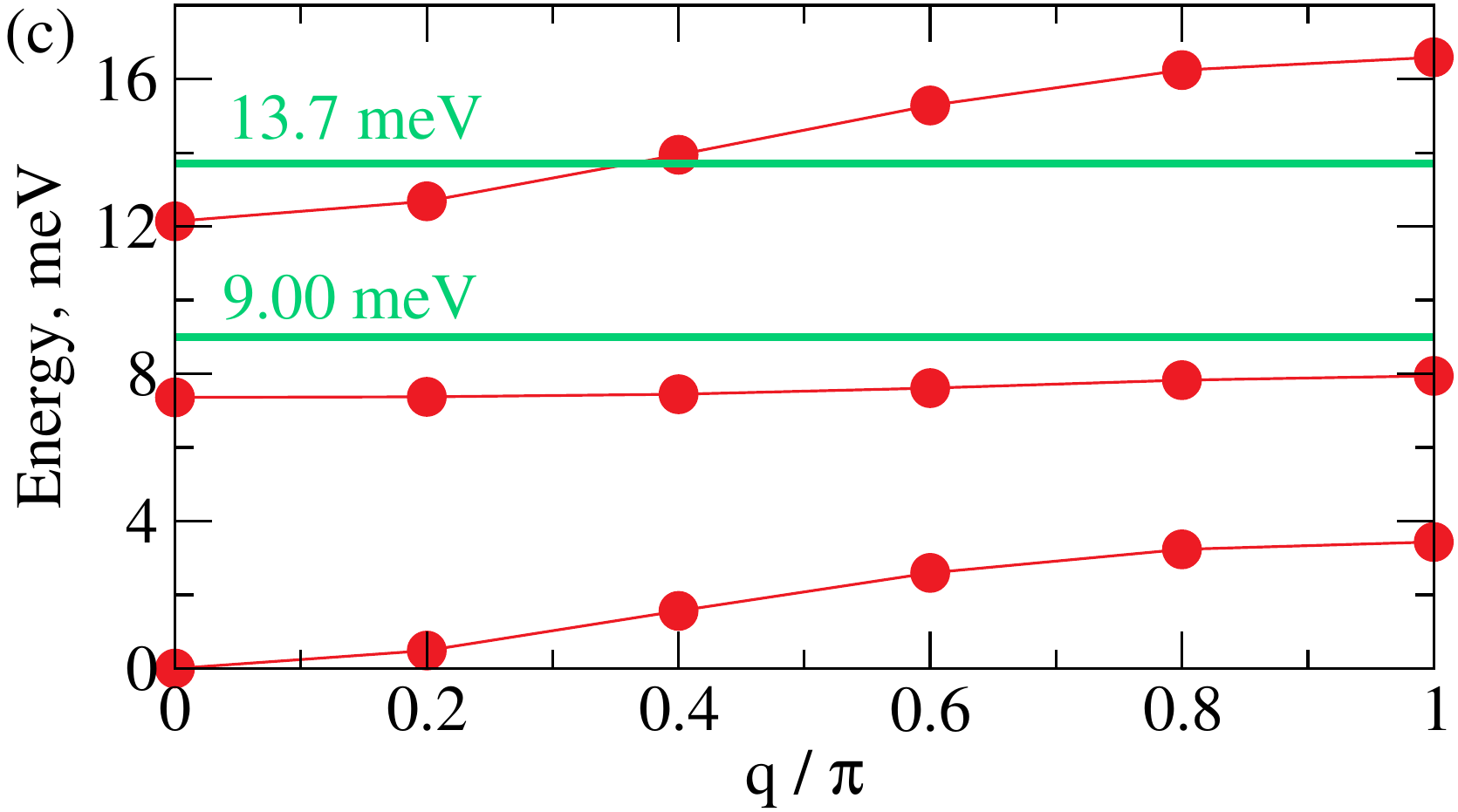}
\caption{(a) Product of the susceptibility $\chi$ 
by the temperature $T$ as a function of $T$. 
Quantum Monte Carlo (QMC) results for $J=0,0.2,0.4\text{, and }1.0$, 
and experimental data from Ref. \onlinecite{Belik2005} 
for the compound Pb$_3$Cu$_3$(PO$_4$)$_4$; in this case we use 
two sets of values for $J_1$ and $g$: $J_1=74.8 \text{ K}$ and $g=2.09$, 
$J_1=96.7 \text{ K}$ and $g=2.20$. (b) $\chi T$ for Pb$_3$Cu$_3$(PO$_4$)$_4$ 
per mole of trimers and QMC results with the indicated  
$J$ with the respective $J_1$ and $g$. (c) Spin-wave modes for $J=0.2$ and 
$J_1=96.7 \text{ K}$ calculated with exact diagonalization for a trimer 
chain with $L=10$, they correspond to the normalized modes $\omega_0^{(-)}$, $\omega_1^{(-)}$ and 
$\omega^{(+)}$ shown in Fig. \ref{fig:espectro13}. We also indicate the central values 
of two excitations modes of the compound Pb$_3$Cu$_3$(PO$_4$)$_4$: $9.00$ meV and 
$13.7$ meV, as was estimated in Ref. \cite{Matsuda2005} at $T=8\text{ K}$
from neutron scattering experiments.}
\label{fig:suscepMat}
\end{figure}

In Fig. \ref{fig:suscepTneg}, we study the magnetic susceptibility of the trimer chain for 
$J<0$. In Fig. \ref{fig:suscepTneg}(a), we present the susceptibility 
per site of the trimer chain, and of the antiferromagnetic
spin--1/2 linear chain. For simplicity, we define $J_1$ as the superexchange coupling between 
the nearest neighbors in the linear chain. The low-$T$ behavior of $\chi$ for the linear 
chain is given by \cite{Eggert1994}
\begin{equation}
 \frac{\chi(T)}{N g^2\mu_B^2}=\frac{1}{2\pi v}+\frac{1}{4\pi v\ln(T/T_0)},
 \label{eq:chilowt}
\end{equation}
where the spin-wave velocity $v=J_1\pi/2$ is the slope of the dispersion 
relation obtained from the Bethe ansatz and $T_0$ is a non-universal constant.
The logarithmic correction implies that $\chi(T)$ has an infinite slope as $T\rightarrow 0$. 
Notwithstanding, Eq. (\ref{eq:chilowt}) is expected to be universal 
\cite{Eggert1994} under some general conditions. Introducing the expression for  
$v$ into Eq. (\ref{eq:chilowt}), we obtain
\begin{equation}
 \frac{\chi J_1}{N g^2\mu_B^2}=\frac{1}{\pi^2}\left \{1+\frac{1}{2\ln(T/T_0)}\right \},
\end{equation}
so that $\chi(0)J_1=1/\pi^2$. The value of $T_0=\sqrt{\pi/8}\exp(\gamma+1/4)J\approx 2.87J_1$ 
is an exact result \cite{Lukyanov1998}, and was confirmed by thermodynamic 
Bethe ansatz calculations \cite{Klumper2000,Johnston2000}. In Fig. \ref{fig:suscepTneg}(a), we also show that $\chi$ for the trimer chain has 
an increasing slope and approaches the value $1/\pi^2$ as $T\rightarrow 0$, in agreement with Eq. (\ref{eq:chilowt}). In fact, much lower temperatures are required 
to obtain the value of $T_0$. Also, since the value of $\chi(0)J_1$ is near $1/\pi^2$, the data 
shows that the spin-wave velocity for the trimer chain is near $v=J_1 \pi/2$ for $J=-1$.
As evidenced in the inset of Fig. \ref{fig:suscepTneg}(a),  
finite-size effects are not appreciable in both chains,
in the range of temperatures exhibited.

In Fig. \ref{fig:suscepTneg}(b), we present $\chi$ for different values of $J$. 
For $J=0$, decoupled trimers, 
the susceptibility has a paramagnetic behavior with $\chi\sim 1/T$ as $T\rightarrow 0$ [see also Fig. \ref{fig:suscepMat}(a)
for $\chi T$]. For $J<0$, the susceptibility curve exhibits a maximum marking the 
crossover from the low-temperature regime to the high-temperature
decoupled-trimer behavior. This maximum is found in a temperature that increases with decreasing $J$.
In the inset of Fig. \ref{fig:suscepTneg}(b), we present $(J_1\chi)$ normalized by $1/|J|$, so that 
$(J_1\chi)/(1/|J|)=|J|(J_1\chi)=|J_2|\chi$. 
For any value of $J$ ($<0$), the data approach $J_1\chi(0)$ for the antiferromagnetic 
linear chain: $1/\pi^2$. A simple reasoning can be used to explain this behavior. Although the 
spin-wave velocity of the trimer chain can have a complex dependence on 
$J_2$, we expect that $v$ is an increasing function of $|J_2|$. Further, the low-energy magnetic excitations 
shown in Fig. \ref{fig:espectro0} suggest that $v$ is nearly proportional to $|J|=|J_2|$, by fixing $J_1\equiv 1$. 
If we use $v=|J_2|\pi/2$ in Eq. (\ref{eq:chilowt}), we obtain that $|J_2|\chi(0)/N g^2\mu_B^2=1/\pi^2$, as suggested by 
the data in the inset of Fig. \ref{fig:suscepTneg}(b).  

We remark that an antiferromagnetic pattern like that of the linear spin-1/2 chain was reported  in Ref. [\onlinecite{Gu2007}] for the unfrustrated diamond chain, and their data also suggests a finite susceptibility at $T=0$.

In Fig. \ref{fig:suscepMat}(a), we present the magnetic susceptibility per trimer for $J>0$, where the trimer 
chain is found in the ferrimagnetic regime.
One of the signatures of the ferrimagnetic phase is a minimum in 
the $\chi T$ versus $T$ curve at zero field. 
This minimum marks a crossover between the low-temperature ferromagnetic regime, for which $\chi\sim 1/T^2$  \cite{Takahashi1985,Yamada1986}, 
and the high-temperature decoupled-trimer ($J=0$) behavior. Thus, the minimum of $\chi T$ and its 
temperature location decrease with $J$. As discussed above, for $J=0$, the data in Fig. \ref{fig:suscepTneg}(b)
show that $\chi\sim 1/T$ as $T\rightarrow 0$, with $\chi T$ becoming very flat in the low-$T$ regime, as 
shown in Fig. \ref{fig:suscepMat}(a). 
In order to estimate the values of $J_1$, $J_2$, and $g$ 
for the phosphate Pb$_3$Cu$_3$(PO$_4$)$_4$, we  
fit the QMC results to the experimental data for this compound \cite{Belik2005}. 
The value of $J_1$ is estimated by comparing the dimensionless value of the 
temperature, $k_B T/J_1$, with its value in the experimental curve for all values of $J$. Further, 
we compare the value of the minimum of $\chi T$ in the experimental curve and the dimensionless QMC data for all values of $J$, thus obtaining $g$ for each value of $J$. 
Finally, with these values of $g$ and $J_1$, we normalize the experimental curve and compare it with the QMC data for all values of $J$ simulated.
Following this procedure, we have found that the experimental curve is best fitted by $J=0.2$ and $J_1=96.7\text{ K}$,
in the low-$T$ regime, while above the crossover temperature, the best fit occurs for $J=0.4$ and $J_1=74.8\text{ K}$, as shown in Fig. \ref{fig:suscepMat}(a) and in Fig. \ref{fig:suscepMat}(b). The curve for $J=0.4$ 
starts to depart from the experimental one at about $T\sim 25 \text{K}$. The three-dimensional ordering temperature for this 
compound is estimated from specific heat data \cite{Belik2005} to be $T_{3D}\approx 1.3\text{ K}$, and 
does not affect the value of $J$ appreciably. Lastly, we also mention that the enhancement of $J$ above the 
crossover temperature could be associated with lattice vibrations.

The trimer chain compound Pb$_3$Cu$_3$(PO$_4$)$_4$ was also investigated through neutron scattering experiments 
\cite{Matsuda2005}. From this study, two excitation modes were observed at $T= 8\text{ K}$: one at 
energy 9.00 meV, and the other at 13.7 meV, as indicated in Fig. \ref{fig:suscepMat}(c). We also show in this figure the magnon modes $\omega_0^{(-)}$, $\omega_1^{(-)}$, and $\omega^{(+)}$, previously presented in Fig. \ref{fig:espectro13}, for $J=0.2$ and $J_1=96.7\text{ K}=8.34\text{ meV}$. 
We thus see in Fig. \ref{fig:suscepMat}(c) that the optical mode $\omega^{(+)}$ is centered at approximately 
13.7 meV, indicating that it is one of the excitations observed in the neutron scattering experiment \cite{Matsuda2005}. Also, the nearly flat mode $\omega_1^{(-)}$ is centered at 7.7 meV, which departs about 16 \% from the second mode found by neutron scattering. The gapless mode, $\omega_1^{(-)}$, is in a range of energies that was not investigated in detail in Ref. \cite{Matsuda2005}.

\section{Closing of the 1/3 -- plateau for $J<-1$}
\label{sec:gapclosing}

The 1/3 -- plateau width decreases as $J$ decreases in the region 
$J<0$, as shown in the phase diagram, Fig. \ref{fig:diag}(a). 
In this section we investigate the 1/3 -- plateau for $J<-1$
and discuss two possibilities for the gap-closing behavior
in this region. 

First, we notice that for a finite value of $J_1$ and $J_2\rightarrow -\infty$, or 
$1/|J|=J_1/|J_2|\rightarrow 0$, the trimers are decoupled with each trimer having a total spin equal
to 3/2; see Fig. \ref{fig:ham}.  Since in this case the chain is composed of totally uncorrelated spin-3/2 trimers, the total spin is degenerate for $h=0$. Thus, any finite magnetic field saturates the magnetization at 
$1/|J|=0$. We remark that the trimer chain is also decoupled for $J=0$, however in this case with 
the trimers in a spin-1/2 total spin with a gap. Thus, a finite magnetic field, $h=1.5$, is required to put the trimers in the spin-3/2 state and the chain in the fully polarized state, as shown in Figs. \ref{fig:diag} and \ref{fig:espectro13}.

Since $|J_2|$ is the higher energy scale 
for $J<-1$, it is better to consider the energy in units 
of $|J_2|$ instead of $J_1$.
Thus, because $h$ is the magnetic field in units of $J_1/g\mu_B$, 
the magnetic field in units of $|J_2|/g\mu_B$ is $h/|J|=hJ_1/|J_2|$, 
and the plateau width (the gap) $\Delta$ in units of $|J_2|/g\mu_B$ 
is given by $\Delta /|J|$.
\begin{figure}
 \includegraphics*[width=0.4\textwidth]{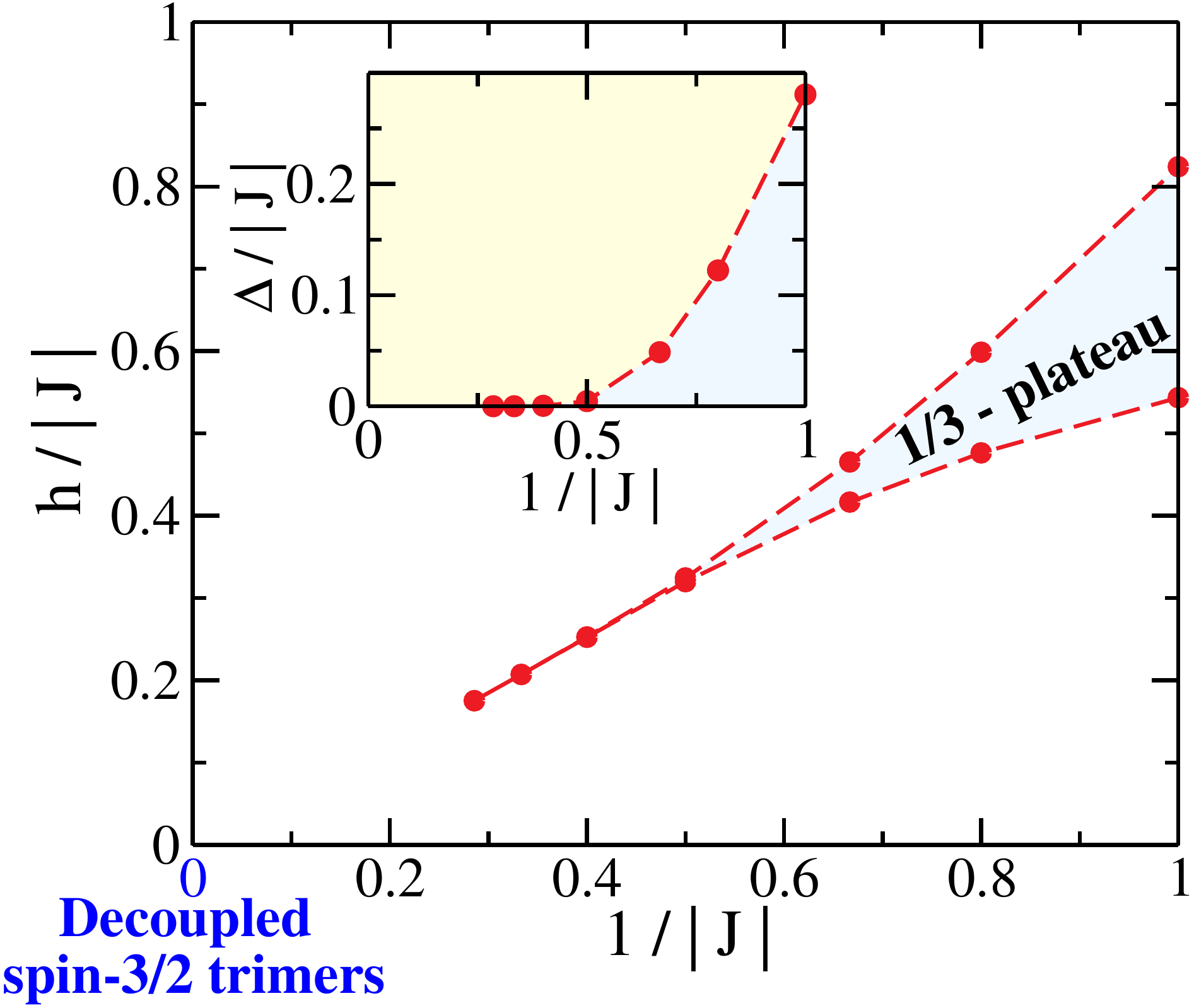}
 \caption{Locus of the magnetization at 1/3 of saturation in the $h/|J|$ (magnetic field 
 in units of $|J_2|/g\mu_B$) versus $1/|J|=J_1/|J_2|$ plane, for $J_2<0$. 
The points are obtained by extrapolating the critical fields 
 from chains of size $L=120,~180,\text{and}~240$. In the inset we present the plateau width 
 $\Delta / |J|$ as a function $1/|J|$. Dashed lines are guides to the eye.} 
\label{fig:diagJneg}
\end{figure}

In Fig. \ref{fig:diagJneg} we show the region where the 1/3 of the saturation 
magnetization, $m_s/3$, is found in the plane $h/|J|$ versus $1/|J|$, and also 
indicate the decoupled-chain limit. The points 
are calculated with DMRG by extrapolating the upper and the lower 
critical fields for $m=m_s/3$ by using their values for $L=120,~180,~\text{and}~240$ 
trimers. We do not show the scale behavior of these critical fields, 
but we mention that it follows a simple straight line as a function of $1/L$ for each value of
$1/|J|$ considered. For the lower value of $1/|J|$ shown, the value of the 
plateau width $\Delta/|J|$ is very tiny, in the limit of our precision. In fact, we present 
in the inset of Fig. \ref{fig:diagJneg} the exponential behavior of the plateau width 
as $1/|J|\rightarrow 0$. 

We can draw two scenarios from the numerical data. In the first one, the plateau width becomes  
exactly null at a value of $1/|J|$ higher than zero. In this case, the transition would be of a 
Kosterlitz-Thouless (KT) type to a gapless Luttinger liquid phase  
with a power-law decay of the transverse spin correlation functions. 
In particular, KT transition points were observed in ferrimagnetic chains \cite{YamamotoPRB99,Montenegro-Filho2020} and branched chains \cite{Verissimo2019,Karlova2019}.
In the other scenario, 
the gap is null only at the decoupled-chain limit, such that a very tiny gap is observed for 
any finite value of $1/|J|$. This behavior is observed in ladder \cite{Barnes1993,Dagotto1996} 
and zigzag models \cite{White1996,Itoi2001} 
as the coupling between the two spin-1/2 chains that compose these systems is reduced to zero. We present
data below that supports the second scenario for the spin trimer chain.
\begin{figure}
 \includegraphics*[width=0.46\textwidth]{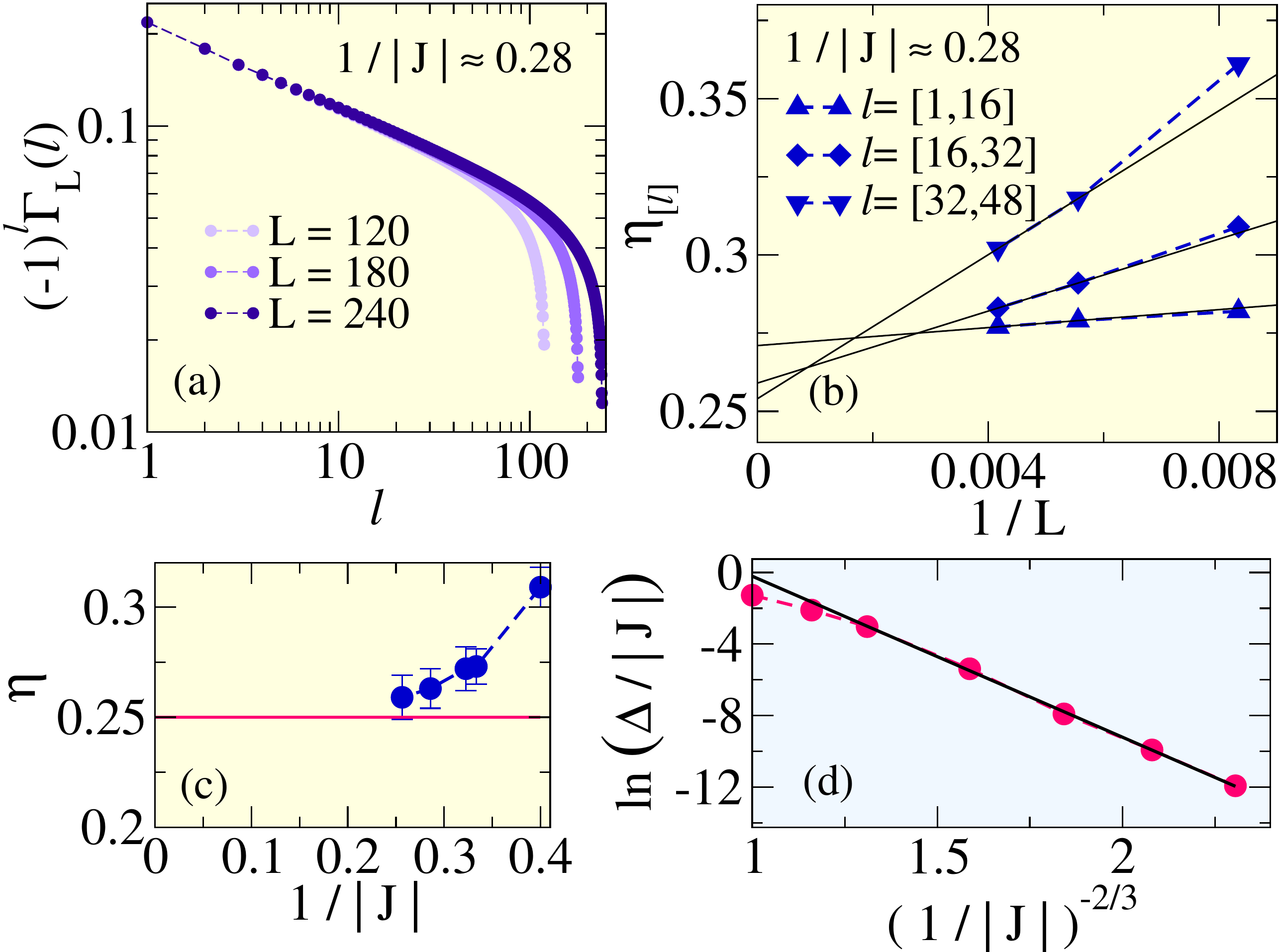}
 \caption{(a) Transverse spin correlation function $\Gamma_L(l)$ between sites in the $A$ sublattice as a function of trimer distance $l$ for the indicated system 
 sizes $L$ and $1/|J|=J_1/|J_2|=1/|-3.5|\approx 0.28$. (b) Exponent $\eta_{[l]}$ calculated from the fitting of $(-1)^l\Gamma_L(l)$ to a function 
 $\sim 1/l^{\eta_{[l]}}$ considering the intervals of $l$ shown in the figure and $1/|J|=1/|-3.5|\approx 0.28$. (c) Exponent $\eta$ as a function of $1/|J|$. (d) Linear behavior of $\ln(\Delta / |J|)$, where $\Delta/|J|$ 
 is the 1/3 -- plateau width in units of $|J_2|/g\mu_B$, as a function of $(1/|J|)^{-2/3}$. The fitting straight 
 line implies that $\frac{\Delta}{|J|} \sim 6600\exp[{-9.0(1/|J|)^{-2/3}}]$ as $|J|\rightarrow \infty$.}
\label{fig:corrJneg}
\end{figure}

To investigate the possibility of a KT transition, we follow a strategy that was successfully used 
in the boson Hubbard model in one dimension \cite{Kuhner} and in an anisotropic ferrimagnetic chain \cite{Montenegro-Filho2020}. In the gapless Luttinger liquid phase, the asymptotic behavior of the transverse spin correlation 
functions is 
\begin{equation}
 \Gamma(l)\sim \frac{1}{l^\eta},
 \label{eq:fitting}
\end{equation}
and $\eta=0.25$ at the KT transition point.
We thus use DMRG to estimate $\eta$ by fitting the transverse spin 
correlation function in finite-size open systems and extrapolating 
the result to $L\rightarrow \infty$. For a chain of size $L$ the transverse spin correlation
function between the spins in the sublattice $A$ is calculated through
\begin{equation}
 \Gamma_L(l)\equiv\langle\langle S^+(i)S^-(i+l)\rangle\rangle_i, 
\end{equation}
where $\langle\langle \ldots \rangle\rangle_i$ indicates the quantum expectation value 
and an average of the correlation over all pairs of $A$ sites
at trimers with a distance $l$ between then, in order to minimize the effects of the
open boundaries. In Fig. \ref{fig:corrJneg}(a) we exemplify the behavior of $\Gamma_L(l)$ 
for $1/|J|=1/|-3.5|\approx 0.28$ for $L = 120, 180,~\text{and}~240$. To estimate $\eta$, we choose three intervals of $l$: [1, 16], [16, 32], and [32, 48], to fit $(-1)^l\Gamma_L(l)$ to Eq. (\ref{eq:fitting}).
Using a linear scale function in $1/L$, we obtain the extrapolated value $\eta_{[l]}$ for each interval. 
This procedure is illustrated in Fig. \ref{fig:corrJneg}(b) for $1/|J|\approx 0.28$. 
The estimated values of $\eta$ are shown in Fig. \ref{fig:corrJneg}(c) as 
a function of $1/|J|$. We define $\eta$ as $(\eta_{[l],\text{max}}+\eta_{[l],\text{min}})/2$, with an error  
$(\eta_{[l],\text{max}}-\eta_{[l],\text{min}})/2$, where $\eta_{[l],\text{max}}$ 
and $\eta_{[l],\text{min}}$ are the maximum and the minimum values of $\eta_{[l]}$, respectively. 
Thus, we did not detect
a point at which $\eta =0.25$ for the values of $1/|J|$ investigated, although $\eta$
approaches 0.25 as $1/|J|\rightarrow 0$. This implies that the KT transition does not occur down  
to the lowest value shown: $1/|-3.9|= 1/3.9 \approx 0.26$. However, the data do not
preclude the possibility of gap closing for $1/|J|\lesssim 0.26$.

Considering the second scenario, that the gap closes exactly at $1/|J|=0$, we have obtained a good 
fitting of our data to the analytical formula \cite{Itoi2001} suggested by field theory to the gap closing
in the case of the zigzag chain near the decoupled two chains from its antiferromagnetic phase. 
The analytical expression \cite{Itoi2001} for the zigzag chain adapted to our model is given by
\begin{equation}
 \frac{\Delta}{|J|}=D\exp{\left[-c\left(\frac{1}{|J|}\right)^{-2/3}\right]},
\end{equation}
such that
\begin{equation}
 \ln\left(\frac{\Delta}{|J|}\right)=\ln{D}-c\left(\frac{1}{|J|}\right)^{-2/3}.
\end{equation}
We show in Fig. \ref{fig:corrJneg}(d) that our gap satisfies this expression for the lowest 
values of $1/|J|$ calculated.   

Thus, with $J_2<0$, down to the lowest values of $1/|J|$, our results 
indicate that the gap closes only at $1/|J|=0$. Further, since the chain is not in 
a critical phase at $1/|J|=0$ for $J_2<0$, we do not have a genuine phase transition
in our case in this scenario. However, a deeper analytical study of the trimer chain, out of the scope 
of this investigation, can put this conclusion in a more firm basis.

\section{Summary}
\label{sec:summary}

We have investigated the rich phase diagram of the spin-1/2 trimer chain 
for $-1\leq J\leq 1$ in the presence and in the absence of a magnetic field. The ground state at zero 
field is ferrimagnetic for $J>0$, in accord with the Lieb-Mattis theorem, and a singlet state for $-1\leq J<0$, a 
range of $J$ for which the Lieb-Mattis theorem does not apply. We have shown that the field-dependent magnetization exhibits the fully polarized plateau, and the 1/3 -- plateau for all values of $J$.
In the 1/3 -- plateau, the unit-cell spins have a ferrimagnetic orientation for
$-0.75\lesssim J<1$, and a ferromagnetic orientation for $-1\leq J\lesssim-0.75$.
For finite-size systems with open boundaries, the 1/3 -- plateau is made of two steps for 
$0.5\lesssim J\leq 1$; likewise, the fully polarized plateau has the same referred feature for $-1\leq J<0$.
The breaking of the plateaus is associated with the magnon occupation of localized 
states at the edges of the chain. We have argued that, for the fully polarized plateau, the 
magnon density at edge states has a purely geometrical origin. In contrast, the edge densities in the 1/3 -- plateau 
have a mixed nature: geometrical and from magnon-magnon interactions. The low-energy bulk magnon 
excitations have three bands: two ferromagnetic -- the first is gapless and dispersive, the second is gapped and nearly flat, for excitations carrying a spin $\Delta S^z=-1$ -- and the third is a dispersive antiferromagnetic mode, carrying a spin $\Delta S^z=+1$. The gapless ferromagnetic mode is unstable for $-1\leq J<0$, in which case the singlet state is the ground state at zero magnetic field. The low-energy excitation from the singlet state, normalized by $|J|$, follows a profile similar to the spinon band of the antiferromagnetic spin-1/2 linear chain. In particular, the minimum energy in the excitation spectrum presents an alternation between $q=0$, for even system sizes excluding multiples of 4, and $q=\pi$, for system sizes that are multiples of 4, as in the spin-1/2 linear chain.

The magnetic susceptibility $\chi$ times the temperature, $\chi T$, 
has a minimum as a function of $T$ in the ferrimagnetic phase, $0 < J\leq 1$.
We determine the values of the coupling constants, $J$ and $J_1$, 
to fit the experimental susceptibility of the ferrimagnetic trimer-chain compound Pb$_3$Cu$_3$(PO$_4$)$_4$. Further, we confirm that the centers of the two gapped magnon bands of the Pb$_3$Cu$_3$(PO$_4$)$_4$ are in accord with the coupling constants estimated from the susceptibility.
For the singlet phase, $-1\leq J<0$, our data evidence that the dimensionless susceptibility per site normalized by $1/|J|$ approaches $1/\pi^2$ as the temperature $T\rightarrow 0$, similarly to  
the spin-1/2 linear chain. Finally, the transverse spin correlation functions and 
the gap behavior for $J\rightarrow-\infty$ points to a closing of the 1/3 -- plateau only 
at $1/|J|=0$ with $J<0$.

\section{ACKNOWLEDGMENTS}
We thank A. A. Belik for sending to us the susceptibility data of the compound 
Pb$_3$Cu$_3$(PO$_4$)$_4$, and useful discussions with M. Matsuda. We also acknowledge the support from Coordenação de Aperfeiçoamento de Pessoal de Nível Superior (CAPES), Conselho Nacional de Desenvolvimento Científico e Tecnológico (CNPq), and Fundação de Amparo à Ciência e Tecnologia do Estado de Pernambuco (FACEPE), Brazilian agencies, including the PRONEX Program which is funded by CNPq and FACEPE, Grant No. APQ-0602-1.05/14. 

\bibliography{Ref}

\begin{thebibliography}{59}%
\makeatletter
\providecommand \@ifxundefined [1]{%
 \@ifx{#1\undefined}
}%
\providecommand \@ifnum [1]{%
 \ifnum #1\expandafter \@firstoftwo
 \else \expandafter \@secondoftwo
 \fi
}%
\providecommand \@ifx [1]{%
 \ifx #1\expandafter \@firstoftwo
 \else \expandafter \@secondoftwo
 \fi
}%
\providecommand \natexlab [1]{#1}%
\providecommand \enquote  [1]{``#1''}%
\providecommand \bibnamefont  [1]{#1}%
\providecommand \bibfnamefont [1]{#1}%
\providecommand \citenamefont [1]{#1}%
\providecommand \href@noop [0]{\@secondoftwo}%
\providecommand \href [0]{\begingroup \@sanitize@url \@href}%
\providecommand \@href[1]{\@@startlink{#1}\@@href}%
\providecommand \@@href[1]{\endgroup#1\@@endlink}%
\providecommand \@sanitize@url [0]{\catcode `\\12\catcode `\$12\catcode
  `\&12\catcode `\#12\catcode `\^12\catcode `\_12\catcode `\%12\relax}%
\providecommand \@@startlink[1]{}%
\providecommand \@@endlink[0]{}%
\providecommand \url  [0]{\begingroup\@sanitize@url \@url }%
\providecommand \@url [1]{\endgroup\@href {#1}{\urlprefix }}%
\providecommand \urlprefix  [0]{URL }%
\providecommand \Eprint [0]{\href }%
\providecommand \doibase [0]{http://dx.doi.org/}%
\providecommand \selectlanguage [0]{\@gobble}%
\providecommand \bibinfo  [0]{\@secondoftwo}%
\providecommand \bibfield  [0]{\@secondoftwo}%
\providecommand \translation [1]{[#1]}%
\providecommand \BibitemOpen [0]{}%
\providecommand \bibitemStop [0]{}%
\providecommand \bibitemNoStop [0]{.\EOS\space}%
\providecommand \EOS [0]{\spacefactor3000\relax}%
\providecommand \BibitemShut  [1]{\csname bibitem#1\endcsname}%
\let\auto@bib@innerbib\@empty
\bibitem [{\citenamefont {Giamarchi}(2004)}]{giamarchi2003quantum}%
  \BibitemOpen
  \bibfield  {author} {\bibinfo {author} {\bibfnamefont {T.}~\bibnamefont
  {Giamarchi}},\ }\href@noop {} {\emph {\bibinfo {title} {Quantum physics in
  one dimension}}}\ (\bibinfo  {publisher} {Clarendon Press, Oxford},\ \bibinfo
  {year} {2004})\BibitemShut {NoStop}%
\bibitem [{\citenamefont {M{\"{u}}ller}\ \emph {et~al.}(1981)\citenamefont
  {M{\"{u}}ller}, \citenamefont {Thomas}, \citenamefont {Beck},\ and\
  \citenamefont {Bonner}}]{Muller1981}%
  \BibitemOpen
  \bibfield  {author} {\bibinfo {author} {\bibfnamefont {G.}~\bibnamefont
  {M{\"{u}}ller}}, \bibinfo {author} {\bibfnamefont {H.}~\bibnamefont
  {Thomas}}, \bibinfo {author} {\bibfnamefont {H.}~\bibnamefont {Beck}}, \ and\
  \bibinfo {author} {\bibfnamefont {J.~C.}\ \bibnamefont {Bonner}},\ }\href
  {\doibase 10.1103/PhysRevB.24.1429} {\bibfield  {journal} {\bibinfo
  {journal} {Physical Review B}\ }\textbf {\bibinfo {volume} {24}},\ \bibinfo
  {pages} {1429} (\bibinfo {year} {1981})}\BibitemShut {NoStop}%
\bibitem [{\citenamefont {Dagotto}\ \emph {et~al.}(1992)\citenamefont
  {Dagotto}, \citenamefont {Riera},\ and\ \citenamefont
  {Scalapino}}]{Dagotto1992}%
  \BibitemOpen
  \bibfield  {author} {\bibinfo {author} {\bibfnamefont {E.}~\bibnamefont
  {Dagotto}}, \bibinfo {author} {\bibfnamefont {J.}~\bibnamefont {Riera}}, \
  and\ \bibinfo {author} {\bibfnamefont {D.}~\bibnamefont {Scalapino}},\ }\href
  {\doibase 10.1103/PhysRevB.45.5744} {\bibfield  {journal} {\bibinfo
  {journal} {Physical Review B}\ }\textbf {\bibinfo {volume} {45}},\ \bibinfo
  {pages} {5744} (\bibinfo {year} {1992})}\BibitemShut {NoStop}%
\bibitem [{\citenamefont {Barnes}\ \emph {et~al.}(1993)\citenamefont {Barnes},
  \citenamefont {Dagotto}, \citenamefont {Riera},\ and\ \citenamefont
  {Swanson}}]{Barnes1993}%
  \BibitemOpen
  \bibfield  {author} {\bibinfo {author} {\bibfnamefont {T.}~\bibnamefont
  {Barnes}}, \bibinfo {author} {\bibfnamefont {E.}~\bibnamefont {Dagotto}},
  \bibinfo {author} {\bibfnamefont {J.}~\bibnamefont {Riera}}, \ and\ \bibinfo
  {author} {\bibfnamefont {E.~S.}\ \bibnamefont {Swanson}},\ }\href {\doibase
  10.1103/PhysRevB.47.3196} {\bibfield  {journal} {\bibinfo  {journal}
  {Physical Review B}\ }\textbf {\bibinfo {volume} {47}},\ \bibinfo {pages}
  {3196} (\bibinfo {year} {1993})}\BibitemShut {NoStop}%
\bibitem [{\citenamefont {Dagotto}\ and\ \citenamefont
  {Rice}(1996)}]{Dagotto1996}%
  \BibitemOpen
  \bibfield  {author} {\bibinfo {author} {\bibfnamefont {E.}~\bibnamefont
  {Dagotto}}\ and\ \bibinfo {author} {\bibfnamefont {T.~M.}\ \bibnamefont
  {Rice}},\ }\href {\doibase 10.1126/science.271.5249.618} {\bibfield
  {journal} {\bibinfo  {journal} {Science}\ }\textbf {\bibinfo {volume}
  {271}},\ \bibinfo {pages} {618} (\bibinfo {year} {1996})}\BibitemShut
  {NoStop}%
\bibitem [{\citenamefont {White}\ \emph {et~al.}(1994)\citenamefont {White},
  \citenamefont {Noack},\ and\ \citenamefont {Scalapino}}]{White1994}%
  \BibitemOpen
  \bibfield  {author} {\bibinfo {author} {\bibfnamefont {S.~R.}\ \bibnamefont
  {White}}, \bibinfo {author} {\bibfnamefont {R.~M.}\ \bibnamefont {Noack}}, \
  and\ \bibinfo {author} {\bibfnamefont {D.~J.}\ \bibnamefont {Scalapino}},\
  }\href {\doibase 10.1103/PhysRevLett.73.886} {\bibfield  {journal} {\bibinfo
  {journal} {Physical Review Letters}\ }\textbf {\bibinfo {volume} {73}},\
  \bibinfo {pages} {886} (\bibinfo {year} {1994})}\BibitemShut {NoStop}%
\bibitem [{\citenamefont {Frischmuth}\ \emph {et~al.}(1996)\citenamefont
  {Frischmuth}, \citenamefont {Ammon},\ and\ \citenamefont
  {Troyer}}]{Frischmuth1996}%
  \BibitemOpen
  \bibfield  {author} {\bibinfo {author} {\bibfnamefont {B.}~\bibnamefont
  {Frischmuth}}, \bibinfo {author} {\bibfnamefont {B.}~\bibnamefont {Ammon}}, \
  and\ \bibinfo {author} {\bibfnamefont {M.}~\bibnamefont {Troyer}},\ }\href
  {\doibase 10.1103/PhysRevB.54.R3714} {\bibfield  {journal} {\bibinfo
  {journal} {Physical Review B}\ }\textbf {\bibinfo {volume} {54}},\ \bibinfo
  {pages} {R3714} (\bibinfo {year} {1996})}\BibitemShut {NoStop}%
\bibitem [{\citenamefont {Lieb}\ and\ \citenamefont
  {Mattis}(1962)}]{LiebMattis}%
  \BibitemOpen
  \bibfield  {author} {\bibinfo {author} {\bibfnamefont {E.}~\bibnamefont
  {Lieb}}\ and\ \bibinfo {author} {\bibfnamefont {D.}~\bibnamefont {Mattis}},\
  }\href {\doibase 10.1063/1.1724276} {\bibfield  {journal} {\bibinfo
  {journal} {Journal of Mathematical Physics}\ }\textbf {\bibinfo {volume}
  {3}},\ \bibinfo {pages} {749} (\bibinfo {year} {1962})}\BibitemShut {NoStop}%
\bibitem [{\citenamefont {Hase}\ \emph {et~al.}(2006)\citenamefont {Hase},
  \citenamefont {Kohno}, \citenamefont {Kitazawa}, \citenamefont {Tsujii},
  \citenamefont {Suzuki}, \citenamefont {Ozawa}, \citenamefont {Kido},
  \citenamefont {Imai},\ and\ \citenamefont {Hu}}]{Hase2006}%
  \BibitemOpen
  \bibfield  {author} {\bibinfo {author} {\bibfnamefont {M.}~\bibnamefont
  {Hase}}, \bibinfo {author} {\bibfnamefont {M.}~\bibnamefont {Kohno}},
  \bibinfo {author} {\bibfnamefont {H.}~\bibnamefont {Kitazawa}}, \bibinfo
  {author} {\bibfnamefont {N.}~\bibnamefont {Tsujii}}, \bibinfo {author}
  {\bibfnamefont {O.}~\bibnamefont {Suzuki}}, \bibinfo {author} {\bibfnamefont
  {K.}~\bibnamefont {Ozawa}}, \bibinfo {author} {\bibfnamefont
  {G.}~\bibnamefont {Kido}}, \bibinfo {author} {\bibfnamefont {M.}~\bibnamefont
  {Imai}}, \ and\ \bibinfo {author} {\bibfnamefont {X.}~\bibnamefont {Hu}},\
  }\href {\doibase 10.1103/PhysRevB.73.104419} {\bibfield  {journal} {\bibinfo
  {journal} {Physical Review B}\ }\textbf {\bibinfo {volume} {73}},\ \bibinfo
  {pages} {104419} (\bibinfo {year} {2006})}\BibitemShut {NoStop}%
\bibitem [{\citenamefont {Gu}\ \emph {et~al.}(2006)\citenamefont {Gu},
  \citenamefont {Su},\ and\ \citenamefont {Gao}}]{Gu2006}%
  \BibitemOpen
  \bibfield  {author} {\bibinfo {author} {\bibfnamefont {B.}~\bibnamefont
  {Gu}}, \bibinfo {author} {\bibfnamefont {G.}~\bibnamefont {Su}}, \ and\
  \bibinfo {author} {\bibfnamefont {S.}~\bibnamefont {Gao}},\ }\href {\doibase
  10.1103/PhysRevB.73.134427} {\bibfield  {journal} {\bibinfo  {journal}
  {Physical Review B}\ }\textbf {\bibinfo {volume} {73}},\ \bibinfo {pages}
  {134427} (\bibinfo {year} {2006})}\BibitemShut {NoStop}%
\bibitem [{\citenamefont {Hase}\ \emph {et~al.}(2007)\citenamefont {Hase},
  \citenamefont {Matsuda}, \citenamefont {Kakurai}, \citenamefont {Ozawa},
  \citenamefont {Kitazawa}, \citenamefont {Tsujii}, \citenamefont
  {D{\"{o}}nni}, \citenamefont {Kohno},\ and\ \citenamefont {Hu}}]{Hase2007}%
  \BibitemOpen
  \bibfield  {author} {\bibinfo {author} {\bibfnamefont {M.}~\bibnamefont
  {Hase}}, \bibinfo {author} {\bibfnamefont {M.}~\bibnamefont {Matsuda}},
  \bibinfo {author} {\bibfnamefont {K.}~\bibnamefont {Kakurai}}, \bibinfo
  {author} {\bibfnamefont {K.}~\bibnamefont {Ozawa}}, \bibinfo {author}
  {\bibfnamefont {H.}~\bibnamefont {Kitazawa}}, \bibinfo {author}
  {\bibfnamefont {N.}~\bibnamefont {Tsujii}}, \bibinfo {author} {\bibfnamefont
  {A.}~\bibnamefont {D{\"{o}}nni}}, \bibinfo {author} {\bibfnamefont
  {M.}~\bibnamefont {Kohno}}, \ and\ \bibinfo {author} {\bibfnamefont
  {X.}~\bibnamefont {Hu}},\ }\href {\doibase 10.1103/PhysRevB.76.064431}
  {\bibfield  {journal} {\bibinfo  {journal} {Physical Review B}\ }\textbf
  {\bibinfo {volume} {76}},\ \bibinfo {pages} {064431} (\bibinfo {year}
  {2007})}\BibitemShut {NoStop}%
\bibitem [{\citenamefont {Cheng}\ \emph {et~al.}(2022)\citenamefont {Cheng},
  \citenamefont {Li}, \citenamefont {Xiong}, \citenamefont {Wu}, \citenamefont
  {Sandvik},\ and\ \citenamefont {Yao}}]{Cheng2022}%
  \BibitemOpen
  \bibfield  {author} {\bibinfo {author} {\bibfnamefont {J.-Q.}\ \bibnamefont
  {Cheng}}, \bibinfo {author} {\bibfnamefont {J.}~\bibnamefont {Li}}, \bibinfo
  {author} {\bibfnamefont {Z.}~\bibnamefont {Xiong}}, \bibinfo {author}
  {\bibfnamefont {H.-Q.}\ \bibnamefont {Wu}}, \bibinfo {author} {\bibfnamefont
  {A.~W.}\ \bibnamefont {Sandvik}}, \ and\ \bibinfo {author} {\bibfnamefont
  {D.-X.}\ \bibnamefont {Yao}},\ }\href {\doibase 10.1038/s41535-021-00416-4}
  {\bibfield  {journal} {\bibinfo  {journal} {npj Quantum Materials}\ }\textbf
  {\bibinfo {volume} {7}},\ \bibinfo {pages} {3} (\bibinfo {year}
  {2022})}\BibitemShut {NoStop}%
\bibitem [{\citenamefont {Verkholyak}\ and\ \citenamefont
  {Stre{\v{c}}ka}(2021)}]{Verkholyak2021}%
  \BibitemOpen
  \bibfield  {author} {\bibinfo {author} {\bibfnamefont {T.}~\bibnamefont
  {Verkholyak}}\ and\ \bibinfo {author} {\bibfnamefont {J.}~\bibnamefont
  {Stre{\v{c}}ka}},\ }\href {\doibase 10.1103/PhysRevB.103.184415} {\bibfield
  {journal} {\bibinfo  {journal} {Physical Review B}\ }\textbf {\bibinfo
  {volume} {103}},\ \bibinfo {pages} {184415} (\bibinfo {year}
  {2021})}\BibitemShut {NoStop}%
\bibitem [{\citenamefont {Hase}\ \emph {et~al.}(2020)\citenamefont {Hase},
  \citenamefont {Pomjakushin}, \citenamefont {Keller}, \citenamefont {Stuhr},
  \citenamefont {D{\"{o}}nni}, \citenamefont {Kohno},\ and\ \citenamefont
  {Tanaka}}]{Hase2020}%
  \BibitemOpen
  \bibfield  {author} {\bibinfo {author} {\bibfnamefont {M.}~\bibnamefont
  {Hase}}, \bibinfo {author} {\bibfnamefont {V.~Y.}\ \bibnamefont
  {Pomjakushin}}, \bibinfo {author} {\bibfnamefont {L.}~\bibnamefont {Keller}},
  \bibinfo {author} {\bibfnamefont {U.}~\bibnamefont {Stuhr}}, \bibinfo
  {author} {\bibfnamefont {A.}~\bibnamefont {D{\"{o}}nni}}, \bibinfo {author}
  {\bibfnamefont {M.}~\bibnamefont {Kohno}}, \ and\ \bibinfo {author}
  {\bibfnamefont {A.}~\bibnamefont {Tanaka}},\ }\href {\doibase
  10.1103/PhysRevB.102.014403} {\bibfield  {journal} {\bibinfo  {journal}
  {Physical Review B}\ }\textbf {\bibinfo {volume} {102}},\ \bibinfo {pages}
  {014403} (\bibinfo {year} {2020})}\BibitemShut {NoStop}%
\bibitem [{\citenamefont {Mac{\^{e}}do}\ \emph {et~al.}(1995)\citenamefont
  {Mac{\^{e}}do}, \citenamefont {dos Santos}, \citenamefont {Coutinho-Filho},\
  and\ \citenamefont {Mac{\^{e}}do}}]{Macedo1995}%
  \BibitemOpen
  \bibfield  {author} {\bibinfo {author} {\bibfnamefont {A.~M.~S.}\
  \bibnamefont {Mac{\^{e}}do}}, \bibinfo {author} {\bibfnamefont {M.~C.}\
  \bibnamefont {dos Santos}}, \bibinfo {author} {\bibfnamefont {M.~D.}\
  \bibnamefont {Coutinho-Filho}}, \ and\ \bibinfo {author} {\bibfnamefont
  {C.~A.}\ \bibnamefont {Mac{\^{e}}do}},\ }\href {\doibase
  10.1103/PhysRevLett.74.1851} {\bibfield  {journal} {\bibinfo  {journal}
  {Physical Review Letters}\ }\textbf {\bibinfo {volume} {74}},\ \bibinfo
  {pages} {1851} (\bibinfo {year} {1995})}\BibitemShut {NoStop}%
\bibitem [{\citenamefont {Raposo}\ and\ \citenamefont
  {Coutinho-Filho}(1997)}]{PRL97Raposo}%
  \BibitemOpen
  \bibfield  {author} {\bibinfo {author} {\bibfnamefont {E.~P.}\ \bibnamefont
  {Raposo}}\ and\ \bibinfo {author} {\bibfnamefont {M.~D.}\ \bibnamefont
  {Coutinho-Filho}},\ }\href {\doibase 10.1103/PhysRevLett.78.4853} {\bibfield
  {journal} {\bibinfo  {journal} {Physical Review Letters}\ }\textbf {\bibinfo
  {volume} {78}},\ \bibinfo {pages} {4853} (\bibinfo {year}
  {1997})}\BibitemShut {NoStop}%
\bibitem [{\citenamefont {Raposo}\ and\ \citenamefont
  {Coutinho-Filho}(1999)}]{PRB99Raposo}%
  \BibitemOpen
  \bibfield  {author} {\bibinfo {author} {\bibfnamefont {E.~P.}\ \bibnamefont
  {Raposo}}\ and\ \bibinfo {author} {\bibfnamefont {M.~D.}\ \bibnamefont
  {Coutinho-Filho}},\ }\href {\doibase 10.1103/PhysRevB.59.14384} {\bibfield
  {journal} {\bibinfo  {journal} {Physical Review B}\ }\textbf {\bibinfo
  {volume} {59}},\ \bibinfo {pages} {14384} (\bibinfo {year}
  {1999})}\BibitemShut {NoStop}%
\bibitem [{\citenamefont {Tian}(1994)}]{Tian94}%
  \BibitemOpen
  \bibfield  {author} {\bibinfo {author} {\bibfnamefont {G.-S.}\ \bibnamefont
  {Tian}},\ }\href {\doibase 10.1088/0305-4470/27/7/012} {\bibfield  {journal}
  {\bibinfo  {journal} {Journal of Physics A}\
  }\textbf {\bibinfo {volume} {27}},\ \bibinfo {pages} {2305} (\bibinfo {year}
  {1994})}\BibitemShut {NoStop}%
\bibitem [{\citenamefont {Yamamoto}\ and\ \citenamefont
  {Ohara}(2007)}]{Yamamoto2007}%
  \BibitemOpen
  \bibfield  {author} {\bibinfo {author} {\bibfnamefont {S.}~\bibnamefont
  {Yamamoto}}\ and\ \bibinfo {author} {\bibfnamefont {J.}~\bibnamefont
  {Ohara}},\ }\href {\doibase 10.1103/PhysRevB.76.014409} {\bibfield  {journal}
  {\bibinfo  {journal} {Physical Review B}\ }\textbf {\bibinfo {volume} {76}},\
  \bibinfo {pages} {014409} (\bibinfo {year} {2007})}\BibitemShut {NoStop}%
\bibitem [{\citenamefont {Takano}\ \emph {et~al.}(1996)\citenamefont {Takano},
  \citenamefont {Kubo},\ and\ \citenamefont {Sakamoto}}]{Takano1996}%
  \BibitemOpen
  \bibfield  {author} {\bibinfo {author} {\bibfnamefont {K.}~\bibnamefont
  {Takano}}, \bibinfo {author} {\bibfnamefont {K.}~\bibnamefont {Kubo}}, \ and\
  \bibinfo {author} {\bibfnamefont {H.}~\bibnamefont {Sakamoto}},\ }\href
  {\doibase 10.1088/0953-8984/8/35/009} {\bibfield  {journal} {\bibinfo
  {journal} {Journal of Physics: Condensed Matter}\ }\textbf {\bibinfo {volume}
  {8}},\ \bibinfo {pages} {6405} (\bibinfo {year} {1996})}\BibitemShut
  {NoStop}%
\bibitem [{\citenamefont {Okamoto}\ \emph {et~al.}(2003)\citenamefont
  {Okamoto}, \citenamefont {Tonegawa},\ and\ \citenamefont
  {Kaburagi}}]{Okamoto2003}%
  \BibitemOpen
  \bibfield  {author} {\bibinfo {author} {\bibfnamefont {K.}~\bibnamefont
  {Okamoto}}, \bibinfo {author} {\bibfnamefont {T.}~\bibnamefont {Tonegawa}}, \
  and\ \bibinfo {author} {\bibfnamefont {M.}~\bibnamefont {Kaburagi}},\ }\href
  {\doibase 10.1088/0953-8984/15/35/307} {\bibfield  {journal} {\bibinfo
  {journal} {Journal of Physics: Condensed Matter}\ }\textbf {\bibinfo {volume}
  {15}},\ \bibinfo {pages} {5979} (\bibinfo {year} {2003})}\BibitemShut
  {NoStop}%
\bibitem [{\citenamefont {Kikuchi}\ \emph {et~al.}(2005)\citenamefont
  {Kikuchi}, \citenamefont {Fujii}, \citenamefont {Chiba}, \citenamefont
  {Mitsudo}, \citenamefont {Idehara}, \citenamefont {Tonegawa}, \citenamefont
  {Okamoto}, \citenamefont {Sakai}, \citenamefont {Kuwai},\ and\ \citenamefont
  {Ohta}}]{Kikuchi2005}%
  \BibitemOpen
  \bibfield  {author} {\bibinfo {author} {\bibfnamefont {H.}~\bibnamefont
  {Kikuchi}}, \bibinfo {author} {\bibfnamefont {Y.}~\bibnamefont {Fujii}},
  \bibinfo {author} {\bibfnamefont {M.}~\bibnamefont {Chiba}}, \bibinfo
  {author} {\bibfnamefont {S.}~\bibnamefont {Mitsudo}}, \bibinfo {author}
  {\bibfnamefont {T.}~\bibnamefont {Idehara}}, \bibinfo {author} {\bibfnamefont
  {T.}~\bibnamefont {Tonegawa}}, \bibinfo {author} {\bibfnamefont
  {K.}~\bibnamefont {Okamoto}}, \bibinfo {author} {\bibfnamefont
  {T.}~\bibnamefont {Sakai}}, \bibinfo {author} {\bibfnamefont
  {T.}~\bibnamefont {Kuwai}}, \ and\ \bibinfo {author} {\bibfnamefont
  {H.}~\bibnamefont {Ohta}},\ }\href {\doibase 10.1103/PhysRevLett.94.227201}
  {\bibfield  {journal} {\bibinfo  {journal} {Physical Review Letters}\
  }\textbf {\bibinfo {volume} {94}},\ \bibinfo {pages} {227201} (\bibinfo
  {year} {2005})}\BibitemShut {NoStop}%
\bibitem [{\citenamefont {Rule}\ \emph {et~al.}(2008)\citenamefont {Rule},
  \citenamefont {Wolter}, \citenamefont {S{\"{u}}llow}, \citenamefont
  {Tennant}, \citenamefont {Br{\"{u}}hl}, \citenamefont {K{\"{o}}hler},
  \citenamefont {Wolf}, \citenamefont {Lang},\ and\ \citenamefont
  {Schreuer}}]{Rule2008}%
  \BibitemOpen
  \bibfield  {author} {\bibinfo {author} {\bibfnamefont {K.~C.}\ \bibnamefont
  {Rule}}, \bibinfo {author} {\bibfnamefont {A.~U.~B.}\ \bibnamefont {Wolter}},
  \bibinfo {author} {\bibfnamefont {S.}~\bibnamefont {S{\"{u}}llow}}, \bibinfo
  {author} {\bibfnamefont {D.~A.}\ \bibnamefont {Tennant}}, \bibinfo {author}
  {\bibfnamefont {A.}~\bibnamefont {Br{\"{u}}hl}}, \bibinfo {author}
  {\bibfnamefont {S.}~\bibnamefont {K{\"{o}}hler}}, \bibinfo {author}
  {\bibfnamefont {B.}~\bibnamefont {Wolf}}, \bibinfo {author} {\bibfnamefont
  {M.}~\bibnamefont {Lang}}, \ and\ \bibinfo {author} {\bibfnamefont
  {J.}~\bibnamefont {Schreuer}},\ }\href {\doibase
  10.1103/PhysRevLett.100.117202} {\bibfield  {journal} {\bibinfo  {journal}
  {Physical Review Letters}\ }\textbf {\bibinfo {volume} {100}},\ \bibinfo
  {pages} {117202} (\bibinfo {year} {2008})}\BibitemShut {NoStop}%
\bibitem [{\citenamefont {Aimo}\ \emph {et~al.}(2009)\citenamefont {Aimo},
  \citenamefont {Kr{\"{a}}mer}, \citenamefont {Klanj{\v{s}}ek}, \citenamefont
  {Horvati{\'{c}}}, \citenamefont {Berthier},\ and\ \citenamefont
  {Kikuchi}}]{Aimo2009}%
  \BibitemOpen
  \bibfield  {author} {\bibinfo {author} {\bibfnamefont {F.}~\bibnamefont
  {Aimo}}, \bibinfo {author} {\bibfnamefont {S.}~\bibnamefont {Kr{\"{a}}mer}},
  \bibinfo {author} {\bibfnamefont {M.}~\bibnamefont {Klanj{\v{s}}ek}},
  \bibinfo {author} {\bibfnamefont {M.}~\bibnamefont {Horvati{\'{c}}}},
  \bibinfo {author} {\bibfnamefont {C.}~\bibnamefont {Berthier}}, \ and\
  \bibinfo {author} {\bibfnamefont {H.}~\bibnamefont {Kikuchi}},\ }\href
  {\doibase 10.1103/PhysRevLett.102.127205} {\bibfield  {journal} {\bibinfo
  {journal} {Physical Review Letters}\ }\textbf {\bibinfo {volume} {102}},\
  \bibinfo {pages} {127205} (\bibinfo {year} {2009})}\BibitemShut {NoStop}%
\bibitem [{\citenamefont {Montenegro-Filho}\ and\ \citenamefont
  {Coutinho-Filho}(2008)}]{Montenegro-Filho2008}%
  \BibitemOpen
  \bibfield  {author} {\bibinfo {author} {\bibfnamefont {R.~R.}\ \bibnamefont
  {Montenegro-Filho}}\ and\ \bibinfo {author} {\bibfnamefont {M.~D.}\
  \bibnamefont {Coutinho-Filho}},\ }\href {\doibase 10.1103/PhysRevB.78.014418}
  {\bibfield  {journal} {\bibinfo  {journal} {Physical Review B}\ }\textbf
  {\bibinfo {volume} {78}},\ \bibinfo {pages} {014418} (\bibinfo {year}
  {2008})}\BibitemShut {NoStop}%
\bibitem [{\citenamefont {do~Nascimento-Junior}\ and\ \citenamefont
  {Montenegro-Filho}(2019)}]{DoNascimento-Junior2019}%
  \BibitemOpen
  \bibfield  {author} {\bibinfo {author} {\bibfnamefont {A.~M.}\ \bibnamefont
  {do~Nascimento-Junior}}\ and\ \bibinfo {author} {\bibfnamefont {R.~R.}\
  \bibnamefont {Montenegro-Filho}},\ }\href {\doibase
  10.1103/PhysRevB.99.064404} {\bibfield  {journal} {\bibinfo  {journal}
  {Physical Review B}\ }\textbf {\bibinfo {volume} {99}},\ \bibinfo {pages}
  {064404} (\bibinfo {year} {2019})}\BibitemShut {NoStop}%
\bibitem [{\citenamefont {Rojas}\ \emph {et~al.}(2021)\citenamefont {Rojas},
  \citenamefont {de~Souza}, \citenamefont {Torrico}, \citenamefont
  {Ver{\'{i}}ssimo}, \citenamefont {Pereira},\ and\ \citenamefont
  {Lyra}}]{Rojas2021}%
  \BibitemOpen
  \bibfield  {author} {\bibinfo {author} {\bibfnamefont {O.}~\bibnamefont
  {Rojas}}, \bibinfo {author} {\bibfnamefont {S.~M.}\ \bibnamefont {de~Souza}},
  \bibinfo {author} {\bibfnamefont {J.}~\bibnamefont {Torrico}}, \bibinfo
  {author} {\bibfnamefont {L.~M.}\ \bibnamefont {Ver{\'{i}}ssimo}}, \bibinfo
  {author} {\bibfnamefont {M.~S.~S.}\ \bibnamefont {Pereira}}, \ and\ \bibinfo
  {author} {\bibfnamefont {M.~L.}\ \bibnamefont {Lyra}},\ }\href {\doibase
  10.1103/PhysRevE.103.042123} {\bibfield  {journal} {\bibinfo  {journal}
  {Physical Review E}\ }\textbf {\bibinfo {volume} {103}},\ \bibinfo {pages}
  {042123} (\bibinfo {year} {2021})}\BibitemShut {NoStop}%
\bibitem [{\citenamefont {Oshikawa}\ \emph {et~al.}(1997)\citenamefont
  {Oshikawa}, \citenamefont {Yamanaka},\ and\ \citenamefont
  {Affleck}}]{OYAPrl97}%
  \BibitemOpen
  \bibfield  {author} {\bibinfo {author} {\bibfnamefont {M.}~\bibnamefont
  {Oshikawa}}, \bibinfo {author} {\bibfnamefont {M.}~\bibnamefont {Yamanaka}},
  \ and\ \bibinfo {author} {\bibfnamefont {I.}~\bibnamefont {Affleck}},\ }\href
  {\doibase 10.1103/PhysRevLett.78.1984} {\bibfield  {journal} {\bibinfo
  {journal} {Physical Review Letters}\ }\textbf {\bibinfo {volume} {78}},\
  \bibinfo {pages} {1984} (\bibinfo {year} {1997})}\BibitemShut {NoStop}%
\bibitem [{\citenamefont {Montenegro-Filho}\ \emph {et~al.}(2020)\citenamefont
  {Montenegro-Filho}, \citenamefont {Matias},\ and\ \citenamefont
  {Coutinho-Filho}}]{Montenegro-Filho2020}%
  \BibitemOpen
  \bibfield  {author} {\bibinfo {author} {\bibfnamefont {R.~R.}\ \bibnamefont
  {Montenegro-Filho}}, \bibinfo {author} {\bibfnamefont {F.~S.}\ \bibnamefont
  {Matias}}, \ and\ \bibinfo {author} {\bibfnamefont {M.~D.}\ \bibnamefont
  {Coutinho-Filho}},\ }\href {\doibase 10.1103/PhysRevB.102.035137} {\bibfield
  {journal} {\bibinfo  {journal} {Physical Review B}\ }\textbf {\bibinfo
  {volume} {102}},\ \bibinfo {pages} {035137} (\bibinfo {year}
  {2020})}\BibitemShut {NoStop}%
\bibitem [{\citenamefont {Furuya}\ and\ \citenamefont
  {Sato}(2021)}]{Furuya2021}%
  \BibitemOpen
  \bibfield  {author} {\bibinfo {author} {\bibfnamefont {S.~C.}\ \bibnamefont
  {Furuya}}\ and\ \bibinfo {author} {\bibfnamefont {M.}~\bibnamefont {Sato}},\
  }\href {\doibase 10.1103/PhysRevB.104.184401} {\bibfield  {journal} {\bibinfo
   {journal} {Physical Review B}\ }\textbf {\bibinfo {volume} {104}},\ \bibinfo
  {pages} {184401} (\bibinfo {year} {2021})}\BibitemShut {NoStop}%
\bibitem [{\citenamefont {White}(1992)}]{White1992}%
  \BibitemOpen
  \bibfield  {author} {\bibinfo {author} {\bibfnamefont {S.~R.}\ \bibnamefont
  {White}},\ }\href {\doibase 10.1103/PhysRevLett.69.2863} {\bibfield
  {journal} {\bibinfo  {journal} {Physical Review Letters}\ }\textbf {\bibinfo
  {volume} {69}},\ \bibinfo {pages} {2863} (\bibinfo {year}
  {1992})}\BibitemShut {NoStop}%
\bibitem [{\citenamefont {White}(1993)}]{White1993}%
  \BibitemOpen
  \bibfield  {author} {\bibinfo {author} {\bibfnamefont {S.~R.}\ \bibnamefont
  {White}},\ }\href {\doibase 10.1103/PhysRevB.48.10345} {\bibfield  {journal}
  {\bibinfo  {journal} {Physical Review B}\ }\textbf {\bibinfo {volume} {48}},\
  \bibinfo {pages} {10345} (\bibinfo {year} {1993})}\BibitemShut {NoStop}%
\bibitem [{\citenamefont {Schollw{\"{o}}ck}(2005)}]{Schollwock2005}%
  \BibitemOpen
  \bibfield  {author} {\bibinfo {author} {\bibfnamefont {U.}~\bibnamefont
  {Schollw{\"{o}}ck}},\ }\href {\doibase 10.1103/RevModPhys.77.259} {\bibfield
  {journal} {\bibinfo  {journal} {Reviews of Modern Physics}\ }\textbf
  {\bibinfo {volume} {77}},\ \bibinfo {pages} {259} (\bibinfo {year}
  {2005})}\BibitemShut {NoStop}%
\bibitem [{\citenamefont {des Cloizeaux}\ and\ \citenamefont
  {Pearson}(1962)}]{DesCloizeaux1962}%
  \BibitemOpen
  \bibfield  {author} {\bibinfo {author} {\bibfnamefont {J.}~\bibnamefont {des
  Cloizeaux}}\ and\ \bibinfo {author} {\bibfnamefont {J.~J.}\ \bibnamefont
  {Pearson}},\ }\href {\doibase 10.1103/PhysRev.128.2131} {\bibfield  {journal}
  {\bibinfo  {journal} {Physical Review}\ }\textbf {\bibinfo {volume} {128}},\
  \bibinfo {pages} {2131} (\bibinfo {year} {1962})}\BibitemShut {NoStop}%
\bibitem [{\citenamefont {Bauer}\ \emph {et~al.}(2011)\citenamefont {Bauer},
  \citenamefont {Carr}, \citenamefont {Evertz}, \citenamefont {Feiguin},
  \citenamefont {Freire}, \citenamefont {Fuchs}, \citenamefont {Gamper},
  \citenamefont {Gukelberger}, \citenamefont {Gull}, \citenamefont {Guertler},
  \citenamefont {Hehn}, \citenamefont {Igarashi}, \citenamefont {Isakov},
  \citenamefont {Koop}, \citenamefont {Ma}, \citenamefont {Mates},
  \citenamefont {Matsuo}, \citenamefont {Parcollet}, \citenamefont
  {Paw{\l}owski}, \citenamefont {Picon}, \citenamefont {Pollet}, \citenamefont
  {Santos}, \citenamefont {Scarola}, \citenamefont {Schollw{\"{o}}ck},
  \citenamefont {Silva}, \citenamefont {Surer}, \citenamefont {Todo},
  \citenamefont {Trebst}, \citenamefont {Troyer}, \citenamefont {Wall},
  \citenamefont {Werner},\ and\ \citenamefont {Wessel}}]{Bauer2011}%
  \BibitemOpen
  \bibfield  {author} {\bibinfo {author} {\bibfnamefont {B.}~\bibnamefont
  {Bauer}}, \bibinfo {author} {\bibfnamefont {L.~D.}\ \bibnamefont {Carr}},
  \bibinfo {author} {\bibfnamefont {H.~G.}\ \bibnamefont {Evertz}}, \bibinfo
  {author} {\bibfnamefont {A.}~\bibnamefont {Feiguin}}, \bibinfo {author}
  {\bibfnamefont {J.}~\bibnamefont {Freire}}, \bibinfo {author} {\bibfnamefont
  {S.}~\bibnamefont {Fuchs}}, \bibinfo {author} {\bibfnamefont
  {L.}~\bibnamefont {Gamper}}, \bibinfo {author} {\bibfnamefont
  {J.}~\bibnamefont {Gukelberger}}, \bibinfo {author} {\bibfnamefont
  {E.}~\bibnamefont {Gull}}, \bibinfo {author} {\bibfnamefont {S.}~\bibnamefont
  {Guertler}}, \bibinfo {author} {\bibfnamefont {A.}~\bibnamefont {Hehn}},
  \bibinfo {author} {\bibfnamefont {R.}~\bibnamefont {Igarashi}}, \bibinfo
  {author} {\bibfnamefont {S.~V.}\ \bibnamefont {Isakov}}, \bibinfo {author}
  {\bibfnamefont {D.}~\bibnamefont {Koop}}, \bibinfo {author} {\bibfnamefont
  {P.~N.}\ \bibnamefont {Ma}}, \bibinfo {author} {\bibfnamefont
  {P.}~\bibnamefont {Mates}}, \bibinfo {author} {\bibfnamefont
  {H.}~\bibnamefont {Matsuo}}, \bibinfo {author} {\bibfnamefont
  {O.}~\bibnamefont {Parcollet}}, \bibinfo {author} {\bibfnamefont
  {G.}~\bibnamefont {Paw{\l}owski}}, \bibinfo {author} {\bibfnamefont {J.~D.}\
  \bibnamefont {Picon}}, \bibinfo {author} {\bibfnamefont {L.}~\bibnamefont
  {Pollet}}, \bibinfo {author} {\bibfnamefont {E.}~\bibnamefont {Santos}},
  \bibinfo {author} {\bibfnamefont {V.~W.}\ \bibnamefont {Scarola}}, \bibinfo
  {author} {\bibfnamefont {U.}~\bibnamefont {Schollw{\"{o}}ck}}, \bibinfo
  {author} {\bibfnamefont {C.}~\bibnamefont {Silva}}, \bibinfo {author}
  {\bibfnamefont {B.}~\bibnamefont {Surer}}, \bibinfo {author} {\bibfnamefont
  {S.}~\bibnamefont {Todo}}, \bibinfo {author} {\bibfnamefont {S.}~\bibnamefont
  {Trebst}}, \bibinfo {author} {\bibfnamefont {M.}~\bibnamefont {Troyer}},
  \bibinfo {author} {\bibfnamefont {M.~L.}\ \bibnamefont {Wall}}, \bibinfo
  {author} {\bibfnamefont {P.}~\bibnamefont {Werner}}, \ and\ \bibinfo {author}
  {\bibfnamefont {S.}~\bibnamefont {Wessel}},\ }\href {\doibase
  10.1088/1742-5468/2011/05/P05001} {\bibfield  {journal} {\bibinfo  {journal}
  {J. Stat. Mech.: Theory Exp.}\ }\textbf {\bibinfo {volume} {2011}},\ \bibinfo
  {pages} {P05001} (\bibinfo {year} {2011})}\BibitemShut {NoStop}%
\bibitem [{\citenamefont {Montenegro-Filho}\ and\ \citenamefont
  {Coutinho-Filho}(2005)}]{PhysA2005}%
  \BibitemOpen
  \bibfield  {author} {\bibinfo {author} {\bibfnamefont {R.}~\bibnamefont
  {Montenegro-Filho}}\ and\ \bibinfo {author} {\bibfnamefont {M.}~\bibnamefont
  {Coutinho-Filho}},\ }\href {\doibase 10.1016/j.physa.2005.05.060} {\bibfield
  {journal} {\bibinfo  {journal} {Physica A}\ }\textbf {\bibinfo {volume} {357}},\ \bibinfo {pages} {173}
  (\bibinfo {year} {2005})}\BibitemShut {NoStop}%
\bibitem [{\citenamefont {Hu}\ \emph {et~al.}(2014)\citenamefont {Hu},
  \citenamefont {Cheng}, \citenamefont {Xu}, \citenamefont {Luo},\ and\
  \citenamefont {Chen}}]{Hu2014}%
  \BibitemOpen
  \bibfield  {author} {\bibinfo {author} {\bibfnamefont {H.}~\bibnamefont
  {Hu}}, \bibinfo {author} {\bibfnamefont {C.}~\bibnamefont {Cheng}}, \bibinfo
  {author} {\bibfnamefont {Z.}~\bibnamefont {Xu}}, \bibinfo {author}
  {\bibfnamefont {H.-G.}\ \bibnamefont {Luo}}, \ and\ \bibinfo {author}
  {\bibfnamefont {S.}~\bibnamefont {Chen}},\ }\href {\doibase
  10.1103/PhysRevB.90.035150} {\bibfield  {journal} {\bibinfo  {journal}
  {Physical Review B}\ }\textbf {\bibinfo {volume} {90}},\ \bibinfo {pages}
  {035150} (\bibinfo {year} {2014})}\BibitemShut {NoStop}%
\bibitem [{\citenamefont {Hu}\ \emph {et~al.}(2015)\citenamefont {Hu},
  \citenamefont {Cheng}, \citenamefont {Luo},\ and\ \citenamefont
  {Chen}}]{Hu2015}%
  \BibitemOpen
  \bibfield  {author} {\bibinfo {author} {\bibfnamefont {H.-P.}\ \bibnamefont
  {Hu}}, \bibinfo {author} {\bibfnamefont {C.}~\bibnamefont {Cheng}}, \bibinfo
  {author} {\bibfnamefont {H.-G.}\ \bibnamefont {Luo}}, \ and\ \bibinfo
  {author} {\bibfnamefont {S.}~\bibnamefont {Chen}},\ }\href {\doibase
  10.1038/srep08433} {\bibfield  {journal} {\bibinfo  {journal} {Scientific
  Reports}\ }\textbf {\bibinfo {volume} {5}},\ \bibinfo {pages} {8433}
  (\bibinfo {year} {2015})}\BibitemShut {NoStop}%
\bibitem [{\citenamefont {Griffith}\ and\ \citenamefont
  {Continentino}(2018)}]{Griffith2018}%
  \BibitemOpen
  \bibfield  {author} {\bibinfo {author} {\bibfnamefont {M.~A.}\ \bibnamefont
  {Griffith}}\ and\ \bibinfo {author} {\bibfnamefont {M.~A.}\ \bibnamefont
  {Continentino}},\ }\href {\doibase 10.1103/PhysRevE.97.012107} {\bibfield
  {journal} {\bibinfo  {journal} {Physical Review E}\ }\textbf {\bibinfo
  {volume} {97}},\ \bibinfo {pages} {012107} (\bibinfo {year}
  {2018})}\BibitemShut {NoStop}%
\bibitem [{\citenamefont {Rufo}\ \emph {et~al.}(2019)\citenamefont {Rufo},
  \citenamefont {Lopes}, \citenamefont {Continentino},\ and\ \citenamefont
  {Griffith}}]{Rufo2019}%
  \BibitemOpen
  \bibfield  {author} {\bibinfo {author} {\bibfnamefont {S.}~\bibnamefont
  {Rufo}}, \bibinfo {author} {\bibfnamefont {N.}~\bibnamefont {Lopes}},
  \bibinfo {author} {\bibfnamefont {M.~A.}\ \bibnamefont {Continentino}}, \
  and\ \bibinfo {author} {\bibfnamefont {M.~A.~R.}\ \bibnamefont {Griffith}},\
  }\href {\doibase 10.1103/PhysRevB.100.195432} {\bibfield  {journal} {\bibinfo
   {journal} {Physical Review B}\ }\textbf {\bibinfo {volume} {100}},\ \bibinfo
  {pages} {195432} (\bibinfo {year} {2019})}\BibitemShut {NoStop}%
\bibitem [{\citenamefont {{Martinez Alvarez}}\ and\ \citenamefont
  {Coutinho-Filho}(2019)}]{MartinezAlvarez2019}%
  \BibitemOpen
  \bibfield  {author} {\bibinfo {author} {\bibfnamefont {V.~M.}\ \bibnamefont
  {{Martinez Alvarez}}}\ and\ \bibinfo {author} {\bibfnamefont {M.~D.}\
  \bibnamefont {Coutinho-Filho}},\ }\href {\doibase 10.1103/PhysRevA.99.013833}
  {\bibfield  {journal} {\bibinfo  {journal} {Physical Review A}\ }\textbf
  {\bibinfo {volume} {99}},\ \bibinfo {pages} {013833} (\bibinfo {year}
  {2019})}\BibitemShut {NoStop}%
\bibitem [{\citenamefont {Rufo}\ \emph {et~al.}(2021)\citenamefont {Rufo},
  \citenamefont {Griffith}, \citenamefont {Lopes},\ and\ \citenamefont
  {Continentino}}]{Rufo2021}%
  \BibitemOpen
  \bibfield  {author} {\bibinfo {author} {\bibfnamefont {S.}~\bibnamefont
  {Rufo}}, \bibinfo {author} {\bibfnamefont {M.~A.~R.}\ \bibnamefont
  {Griffith}}, \bibinfo {author} {\bibfnamefont {N.}~\bibnamefont {Lopes}}, \
  and\ \bibinfo {author} {\bibfnamefont {M.~A.}\ \bibnamefont {Continentino}},\
  }\href {\doibase 10.1038/s41598-021-01888-x} {\bibfield  {journal} {\bibinfo
  {journal} {Scientific Reports}\ }\textbf {\bibinfo {volume} {11}},\ \bibinfo
  {pages} {22524} (\bibinfo {year} {2021})}\BibitemShut {NoStop}%
\bibitem [{\citenamefont {Watanabe}\ \emph {et~al.}(2021)\citenamefont
  {Watanabe}, \citenamefont {Kato}, \citenamefont {Po},\ and\ \citenamefont
  {Motome}}]{Watanabe2021}%
  \BibitemOpen
  \bibfield  {author} {\bibinfo {author} {\bibfnamefont {H.}~\bibnamefont
  {Watanabe}}, \bibinfo {author} {\bibfnamefont {Y.}~\bibnamefont {Kato}},
  \bibinfo {author} {\bibfnamefont {H.~C.}\ \bibnamefont {Po}}, \ and\ \bibinfo
  {author} {\bibfnamefont {Y.}~\bibnamefont {Motome}},\ }\href {\doibase
  10.1103/PhysRevB.103.134430} {\bibfield  {journal} {\bibinfo  {journal}
  {Physical Review B}\ }\textbf {\bibinfo {volume} {103}},\ \bibinfo {pages}
  {134430} (\bibinfo {year} {2021})}\BibitemShut {NoStop}%
\bibitem [{\citenamefont {da~Silva}\ and\ \citenamefont
  {Montenegro-Filho}(2021)}]{DaSilva2021}%
  \BibitemOpen
  \bibfield  {author} {\bibinfo {author} {\bibfnamefont {W.~M.}\ \bibnamefont
  {da~Silva}}\ and\ \bibinfo {author} {\bibfnamefont {R.~R.}\ \bibnamefont
  {Montenegro-Filho}},\ }\href {\doibase 10.1103/PhysRevB.103.054432}
  {\bibfield  {journal} {\bibinfo  {journal} {Physical Review B}\ }\textbf
  {\bibinfo {volume} {103}},\ \bibinfo {pages} {054432} (\bibinfo {year}
  {2021})}\BibitemShut {NoStop}%
\bibitem [{\citenamefont {Belik}\ \emph {et~al.}(2005)\citenamefont {Belik},
  \citenamefont {Matsuo}, \citenamefont {Azuma}, \citenamefont {Kindo},\ and\
  \citenamefont {Takano}}]{Belik2005}%
  \BibitemOpen
  \bibfield  {author} {\bibinfo {author} {\bibfnamefont {A.~A.}\ \bibnamefont
  {Belik}}, \bibinfo {author} {\bibfnamefont {A.}~\bibnamefont {Matsuo}},
  \bibinfo {author} {\bibfnamefont {M.}~\bibnamefont {Azuma}}, \bibinfo
  {author} {\bibfnamefont {K.}~\bibnamefont {Kindo}}, \ and\ \bibinfo {author}
  {\bibfnamefont {M.}~\bibnamefont {Takano}},\ }\href {\doibase
  10.1016/j.jssc.2004.12.020} {\bibfield  {journal} {\bibinfo  {journal}
  {Journal of Solid State Chemistry}\ }\textbf {\bibinfo {volume} {178}},\
  \bibinfo {pages} {709} (\bibinfo {year} {2005})}\BibitemShut {NoStop}%
\bibitem [{\citenamefont {Matsuda}\ \emph {et~al.}(2005)\citenamefont
  {Matsuda}, \citenamefont {Kakurai}, \citenamefont {Belik}, \citenamefont
  {Azuma}, \citenamefont {Takano},\ and\ \citenamefont {Fujita}}]{Matsuda2005}%
  \BibitemOpen
  \bibfield  {author} {\bibinfo {author} {\bibfnamefont {M.}~\bibnamefont
  {Matsuda}}, \bibinfo {author} {\bibfnamefont {K.}~\bibnamefont {Kakurai}},
  \bibinfo {author} {\bibfnamefont {A.~A.}\ \bibnamefont {Belik}}, \bibinfo
  {author} {\bibfnamefont {M.}~\bibnamefont {Azuma}}, \bibinfo {author}
  {\bibfnamefont {M.}~\bibnamefont {Takano}}, \ and\ \bibinfo {author}
  {\bibfnamefont {M.}~\bibnamefont {Fujita}},\ }\href {\doibase
  10.1103/PhysRevB.71.144411} {\bibfield  {journal} {\bibinfo  {journal}
  {Physical Review B}\ }\textbf {\bibinfo {volume} {71}},\ \bibinfo {pages}
  {144411} (\bibinfo {year} {2005})}\BibitemShut {NoStop}%
\bibitem [{\citenamefont {Eggert}\ \emph {et~al.}(1994)\citenamefont {Eggert},
  \citenamefont {Affleck},\ and\ \citenamefont {Takahashi}}]{Eggert1994}%
  \BibitemOpen
  \bibfield  {author} {\bibinfo {author} {\bibfnamefont {S.}~\bibnamefont
  {Eggert}}, \bibinfo {author} {\bibfnamefont {I.}~\bibnamefont {Affleck}}, \
  and\ \bibinfo {author} {\bibfnamefont {M.}~\bibnamefont {Takahashi}},\ }\href
  {\doibase 10.1103/PhysRevLett.73.332} {\bibfield  {journal} {\bibinfo
  {journal} {Physical Review Letters}\ }\textbf {\bibinfo {volume} {73}},\
  \bibinfo {pages} {332} (\bibinfo {year} {1994})}\BibitemShut {NoStop}%
\bibitem [{\citenamefont {Lukyanov}(1998)}]{Lukyanov1998}%
  \BibitemOpen
  \bibfield  {author} {\bibinfo {author} {\bibfnamefont {S.}~\bibnamefont
  {Lukyanov}},\ }\href {\doibase 10.1016/S0550-3213(98)00249-1} {\bibfield
  {journal} {\bibinfo  {journal} {Nuclear Physics B}\ }\textbf {\bibinfo
  {volume} {522}},\ \bibinfo {pages} {533} (\bibinfo {year}
  {1998})}\BibitemShut {NoStop}%
\bibitem [{\citenamefont {Kl{\"{u}}mper}\ and\ \citenamefont
  {Johnston}(2000)}]{Klumper2000}%
  \BibitemOpen
  \bibfield  {author} {\bibinfo {author} {\bibfnamefont {A.}~\bibnamefont
  {Kl{\"{u}}mper}}\ and\ \bibinfo {author} {\bibfnamefont {D.~C.}\ \bibnamefont
  {Johnston}},\ }\href {\doibase 10.1103/PhysRevLett.84.4701} {\bibfield
  {journal} {\bibinfo  {journal} {Physical Review Letters}\ }\textbf {\bibinfo
  {volume} {84}},\ \bibinfo {pages} {4701} (\bibinfo {year}
  {2000})}\BibitemShut {NoStop}%
\bibitem [{\citenamefont {Johnston}\ \emph {et~al.}(2000)\citenamefont
  {Johnston}, \citenamefont {Kremer},\ and\ \citenamefont
  {Troyer}}]{Johnston2000}%
  \BibitemOpen
  \bibfield  {author} {\bibinfo {author} {\bibfnamefont {D.}~\bibnamefont
  {Johnston}}, \bibinfo {author} {\bibfnamefont {R.}~\bibnamefont {Kremer}}, \
  and\ \bibinfo {author} {\bibfnamefont {M.}~\bibnamefont {Troyer}},\ }\href
  {\doibase 10.1103/PhysRevB.61.9558} {\bibfield  {journal} {\bibinfo
  {journal} {Physical Review B}\ }\textbf {\bibinfo {volume} {61}},\ \bibinfo
  {pages} {9558} (\bibinfo {year} {2000})}\BibitemShut {NoStop}%
\bibitem [{\citenamefont {Gu}\ and\ \citenamefont {Su}(2007)}]{Gu2007}%
  \BibitemOpen
  \bibfield  {author} {\bibinfo {author} {\bibfnamefont {B.}~\bibnamefont
  {Gu}}\ and\ \bibinfo {author} {\bibfnamefont {G.}~\bibnamefont {Su}},\ }\href
  {\doibase 10.1103/PhysRevB.75.174437} {\bibfield  {journal} {\bibinfo
  {journal} {Physical Review B}\ }\textbf {\bibinfo {volume} {75}},\ \bibinfo
  {pages} {174437} (\bibinfo {year} {2007})}\BibitemShut {NoStop}%
\bibitem [{\citenamefont {Takahashi}\ and\ \citenamefont
  {Yamada}(1985)}]{Takahashi1985}%
  \BibitemOpen
  \bibfield  {author} {\bibinfo {author} {\bibfnamefont {M.}~\bibnamefont
  {Takahashi}}\ and\ \bibinfo {author} {\bibfnamefont {M.}~\bibnamefont
  {Yamada}},\ }\href {\doibase 10.1143/JPSJ.54.2808} {\bibfield  {journal}
  {\bibinfo  {journal} {Journal of the Physical Society of Japan}\ }\textbf
  {\bibinfo {volume} {54}},\ \bibinfo {pages} {2808} (\bibinfo {year}
  {1985})}\BibitemShut {NoStop}%
\bibitem [{\citenamefont {Yamada}\ and\ \citenamefont
  {Takahashi}(1986)}]{Yamada1986}%
  \BibitemOpen
  \bibfield  {author} {\bibinfo {author} {\bibfnamefont {M.}~\bibnamefont
  {Yamada}}\ and\ \bibinfo {author} {\bibfnamefont {M.}~\bibnamefont
  {Takahashi}},\ }\href {\doibase 10.1143/JPSJ.55.2024} {\bibfield  {journal}
  {\bibinfo  {journal} {Journal of the Physical Society of Japan}\ }\textbf
  {\bibinfo {volume} {55}},\ \bibinfo {pages} {2024} (\bibinfo {year}
  {1986})}\BibitemShut {NoStop}%
\bibitem [{\citenamefont {Sakai}\ and\ \citenamefont
  {Yamamoto}(1999)}]{YamamotoPRB99}%
  \BibitemOpen
  \bibfield  {author} {\bibinfo {author} {\bibfnamefont {T.}~\bibnamefont
  {Sakai}}\ and\ \bibinfo {author} {\bibfnamefont {S.}~\bibnamefont
  {Yamamoto}},\ }\href {\doibase 10.1103/PhysRevB.60.4053} {\bibfield
  {journal} {\bibinfo  {journal} {Physical Review B}\ }\textbf {\bibinfo
  {volume} {60}},\ \bibinfo {pages} {4053} (\bibinfo {year}
  {1999})}\BibitemShut {NoStop}%
\bibitem [{\citenamefont {Veríssimo}\ \emph {et~al.}(2019)\citenamefont
  {Veríssimo}, \citenamefont {Pereira}, \citenamefont {Strečka},\ and\
  \citenamefont {Lyra}}]{Verissimo2019}%
  \BibitemOpen
  \bibfield  {author} {\bibinfo {author} {\bibfnamefont {L.~M.}\ \bibnamefont
  {Veríssimo}}, \bibinfo {author} {\bibfnamefont {M.~S.~S.}\ \bibnamefont
  {Pereira}}, \bibinfo {author} {\bibfnamefont {J.}~\bibnamefont {Strečka}}, \
  and\ \bibinfo {author} {\bibfnamefont {M.~L.}\ \bibnamefont {Lyra}},\ }\href
  {\doibase 10.1103/PhysRevB.99.134408} {\bibfield  {journal} {\bibinfo
  {journal} {Physical Review B}\ }\textbf {\bibinfo {volume} {99}},\ \bibinfo
  {pages} {134408} (\bibinfo {year} {2019})}\BibitemShut {NoStop}%
\bibitem [{\citenamefont {Karl'ová}\ \emph {et~al.}(2019)\citenamefont
  {Karl'ová}, \citenamefont {Strečka},\ and\ \citenamefont
  {Lyra}}]{Karlova2019}%
  \BibitemOpen
  \bibfield  {author} {\bibinfo {author} {\bibfnamefont {K.}~\bibnamefont
  {Karl'ová}}, \bibinfo {author} {\bibfnamefont {J.}~\bibnamefont {Strečka}},
  \ and\ \bibinfo {author} {\bibfnamefont {M.~L.}\ \bibnamefont {Lyra}},\
  }\href {\doibase 10.1103/PhysRevE.100.042127} {\bibfield  {journal} {\bibinfo
   {journal} {Physical Review E}\ }\textbf {\bibinfo {volume} {100}},\ \bibinfo
  {pages} {042127} (\bibinfo {year} {2019})}\BibitemShut {NoStop}%
\bibitem [{\citenamefont {White}\ and\ \citenamefont
  {Affleck}(1996)}]{White1996}%
  \BibitemOpen
  \bibfield  {author} {\bibinfo {author} {\bibfnamefont {S.~R.}\ \bibnamefont
  {White}}\ and\ \bibinfo {author} {\bibfnamefont {I.}~\bibnamefont
  {Affleck}},\ }\href {\doibase 10.1103/PhysRevB.54.9862} {\bibfield  {journal}
  {\bibinfo  {journal} {Physical Review B}\ }\textbf {\bibinfo {volume} {54}},\
  \bibinfo {pages} {9862} (\bibinfo {year} {1996})}\BibitemShut {NoStop}%
\bibitem [{\citenamefont {Itoi}\ and\ \citenamefont {Qin}(2001)}]{Itoi2001}%
  \BibitemOpen
  \bibfield  {author} {\bibinfo {author} {\bibfnamefont {C.}~\bibnamefont
  {Itoi}}\ and\ \bibinfo {author} {\bibfnamefont {S.}~\bibnamefont {Qin}},\
  }\href {\doibase 10.1103/PhysRevB.63.224423} {\bibfield  {journal} {\bibinfo
  {journal} {Physical Review B}\ }\textbf {\bibinfo {volume} {63}},\ \bibinfo
  {pages} {224423} (\bibinfo {year} {2001})}\BibitemShut {NoStop}%
\bibitem [{\citenamefont {Kühner}\ \emph {et~al.}(2000)\citenamefont
  {Kühner}, \citenamefont {White},\ and\ \citenamefont {Monien}}]{Kuhner}%
  \BibitemOpen
  \bibfield  {author} {\bibinfo {author} {\bibfnamefont {T.~D.}\ \bibnamefont
  {Kühner}}, \bibinfo {author} {\bibfnamefont {S.~R.}\ \bibnamefont {White}},
  \ and\ \bibinfo {author} {\bibfnamefont {H.}~\bibnamefont {Monien}},\ }\href
  {\doibase 10.1103/PhysRevB.61.12474} {\bibfield  {journal} {\bibinfo
  {journal} {Physical Review B}\ }\textbf {\bibinfo {volume} {61}},\ \bibinfo
  {pages} {12474} (\bibinfo {year} {2000})}\BibitemShut {NoStop}%
\end{thebibliography}%
\end{document}